\documentclass[]{aa}
\usepackage{amsmath}
\usepackage{graphicx,color}
\usepackage{natbib}
\usepackage{microtype}
\usepackage{url}
\usepackage[varg]{txfonts}
\usepackage{threeparttable}
\usepackage{booktabs, makecell}
\usepackage{longtable}
\usepackage[utf8]{inputenc}
\usepackage{amsmath}
\usepackage{verbatim}
\usepackage{txfonts}

\usepackage[]{hyperref}%
\def\gai{\textit{Gaia}\xspace}

\def\ros{\textit{ROSAT}\xspace}

\def\tes{\textit{TESS}\xspace}
\def\ero{\textit{eROSITA}\xspace}
\def\xmmn{\textit{XMM-Newton}\xspace}

\def\srg{{SRGE0758}\xspace}

\newcommand\fergs{\ensuremath{\mathrm{erg}\,\mathrm{cm}^{-2}\,\mathrm{s}^{-1}}\xspace}

\begin{document}

\title{Serendipitous discovery of the magnetic cataclysmic variable SRGE J075818-612027}
  
\author{Samet Ok \inst{1,2}
           \and Georg Lamer \inst{1}        
            \and Axel Schwope\inst{1}
             \and David A. H. Buckley \inst{3,4,5}
              \and Jaco Brink \inst{3,4}
               \and Jan Kurpas \inst{1}
               \and Dus\'an Tub\'in \inst{1}
                \and Iris Traulsen \inst{1}
          }
\institute{Leibniz-Institut für Astrophysik Potsdam (AIP), An der Sternwarte 16, 14482 Potsdam, Germany\\
\email{sok@aip.de}
\and
Department of Astronomy \& Space Sciences, Faculty of Science, University of Ege, 35100 Bornova, Izmir, Turkey
\and
South African Astronomical Observatory, PO Box 9, Observatory Road, Observatory 7935, Cape Town, South Africa
\and
Department of Astronomy, University of Cape Town, Private Bag X3, Rondebosch 7701, South Africa
\and
Department of Physics, University of the Free State, PO Box 339, Bloemfontein 9300, South Africa
}

\date{Received ...; Accepted ...}
\keywords{cataclysmic variable stars -- binary stars --
                x-rays --  individual: SRGE J075818-612027 --  stars: fundamental parameters}
                
\abstract{We report the discovery of SRGE J075818-612027, a deep stream-eclipsing magnetic cataclysmic variable found serendipitously in \textit{SRG}/\ero CalPV observations of the open cluster NGC~2516 as an unrelated X-ray source. An X-ray timing and spectral analysis of the \ero data is presented and supplemented by an analysis of \tes photometry and SALT spectroscopy. X-ray photometry reveals two pronounced dips repeating with a period of $106.144(1)$\,min. The 14-month \tes data reveal the same unique period. A low-resolution identification spectrum obtained with SALT displays hydrogen Balmer emission lines on a fairly blue continuum. The spectrum and the stability of the photometric signal led to the classification of the new object as a polar type cataclysmic variable. In this picture the dips in the X-ray light curve are explained by absorption in the intervening accretion stream and by a self-eclipse of the main accretion region. The object displays large magnitude differences on long (months) timescales both at optical and X-ray wavelengths, being interpreted as high and low states and thus supporting the identification as a polar. 
The bright phase X-ray spectrum can be reflected with single temperature thermal emission with 9.7 keV and bolometric X-ray luminosity $L_{\rm X} \simeq 8\times 10^{32}$\,erg s$^{-1}$ at a distance of about 2.7\,kpc. It lacks the pronounced soft X-ray emission component prominently found in \ros-discovered polars.}

\maketitle
%

\section{Introduction}

AM Her systems or polars are interacting close binaries that consist of an accreting white dwarf (WD, called primary) and a Roche-lobe filling donor star, typically on the main sequence, often referred to as the secondary. They are magnetic cataclysmic binaries or cataclysmic variables (CVs). Hence, interaction takes place via Roche lobe overflow to the WD. The WDs in these systems have strong magnetic fields, $B> 10$\, MG, that keeps both stars in synchronous rotation, suppresses an accretion disk, guides accreted matter to the polar regions of the WD, where its potential energy finally is released through radiation from the cooling shocked plasma at X-ray wavelengths and via cyclotron radiation at optical and neighboring wavelength regimes  \citep{warner+95}. 

Polars were found as soft X-ray emitters with the Einstein satellite and with \textit{Röntgensatellit} (\ros). Soft X-ray emission of reprocessed origin or from deeply buried filamentary accretion shocks \citep{lamb+79, frank+88} was regarded one of the observational hallmarks of the class. 

Interestingly, the polar survey conducted by \cite{ramsay+04} with the X-ray Multi-Mirror Mission (\xmmn) and all serendipitous discoveries made with \xmmn since then \citep{vogel+08, ramsay+09, webb+18, schwope+20} did not reveal any further evidence for soft X-ray emission from a large number of magnetic CVs. Also, the first \ero-discovered polar did not show pronounced soft X-ray emission \citep{schwope+22}. Meanwhile about 150 magnetic CVs are known that belong to the polar subclass of the CV population \citep{ritter+03}.

Polars are in the last steps of the binary star evolution and typically have short orbital periods (P$_{orb}$< 2 hours) \citep{ritter+03,knigge+06}. The magnetic torques keep the spin of the WDs synchronized with the orbit so that for most polars $P_{\rm orbit} = P_{\rm spin}$. 

Accordingly, accretion-related emission is modulated on this period. Polars may show alternating high and low accretion states due to cessation of the mass flow from the donor to the WD. Such accretion rate changes thought to be related to the magnetic activity of the secondary star \citep{livio+94}. The optical brightness of the prototype polar, AM Herculis, which has more than a century of optical observations, differs by 2.5 mag between high and low states \citep{hessman+00, wu+08, schwope+20}.

The extended Roentgen Survey with an Imaging Telescope Array (\ero) instrument \citep{predehl+21} on  board  the Spektrum-Roentgen-Gamma  spacecraft \citep[SRG;][]{sunyaev+21} was launched in July 2019 and brought into a wide halo orbit around the Sun-Earth $L_2$ point. Its main aim is to perform several X-ray all-sky surveys (called eRASS), each lasting six month, in the energy range of $0.2 - 10$\,keV. 

Prior to the main survey a comprehensive CalPV (Calibration and Performance Verification) phase of the mission took place, one of the targets was the galactic open cluster NGC\,2516 (also known as Caldwell 96), which was observed to calibrate the boresight and plate scale. The analysis of the data on the cluster member stars will be reported in (Fritzewski et al., in prep.). Here we report on the serendipitous discovery of a new magnetic cataclysmic variable in the field of NGC\,2516, SRGE J075818-612027 or hereafter \srg. The object could be identified initially based on its regular variability pattern which became obvious by a visual inspection of the data cube made of the distribution of events  in the space spanned by the $x,y,$ and time coordinates. The object was also reported as  variable in Gaia DR3 catalogue with a variability flag that indicates long-term photometric variations \citep[see][and references therein]{eyer+22}. 

We detail the X-ray observations obtained by \ero with the comprehensive photometric Transiting Exoplanet Survey Satellite (\tes) data and descriptive spectral observation in Sec.~\ref{s:obs}. We present the results of the basic features of the system by considering the X-ray and optical data together in Sec.~\ref{s:ana}.

\begin{figure}
\resizebox{\hsize}{!}{\includegraphics[width=\columnwidth]{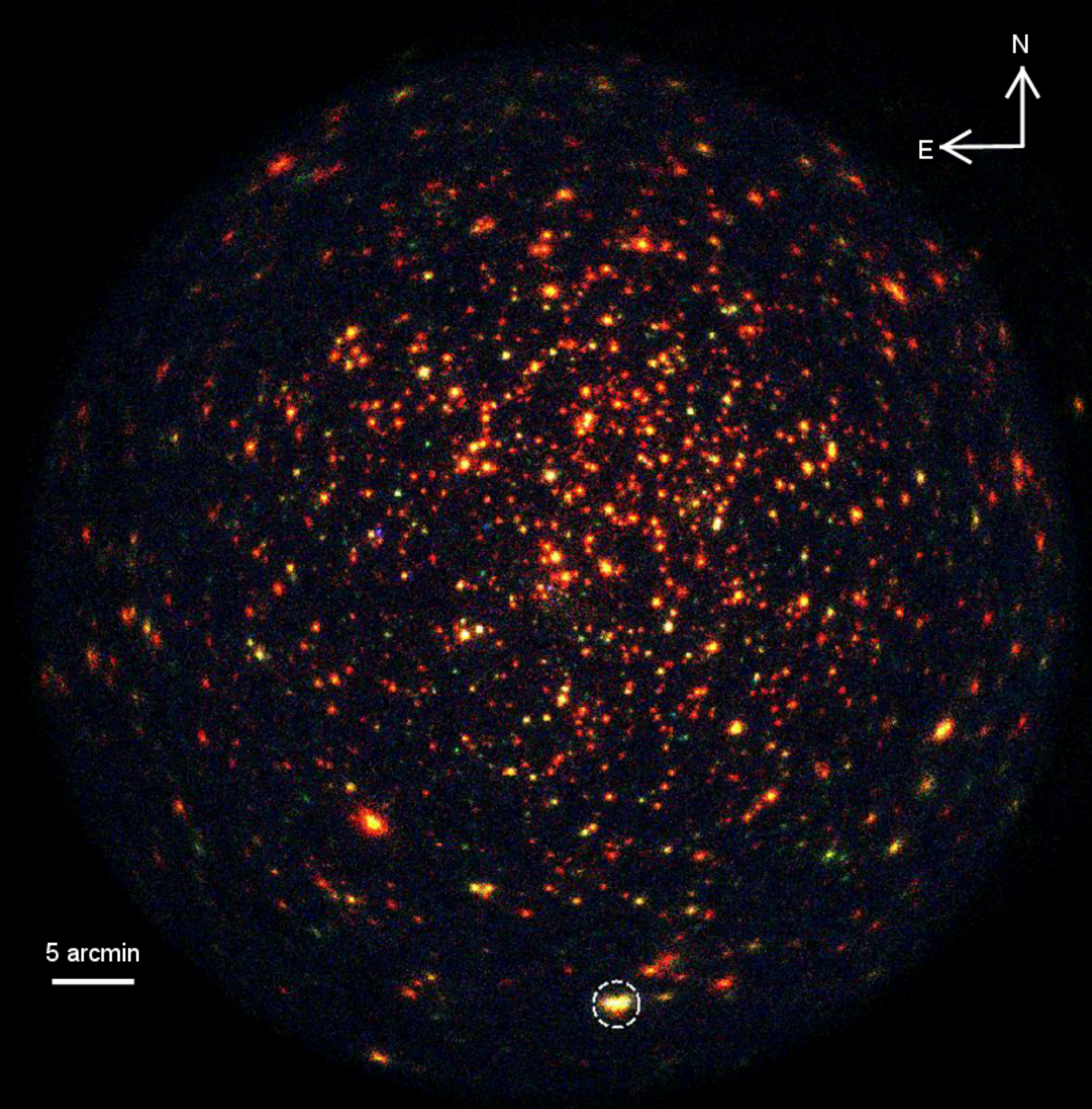}}
\caption{X-ray image of the NGC~2516 obtained by \ero. The white dashed circle marks the position of \srg. False colors of the image were made using photon energies between $0.2 - 1.1$ keV, $1.1 - 2.3$ keV and $2.3 - 5$ keV for the red, green,  and blue channels, respectively. The image displays an area of 1$\times$1 degree.
\label{f:erorgb}}
\end{figure}

\begin{figure}
\resizebox{\hsize}{!}{\includegraphics[width=\columnwidth]{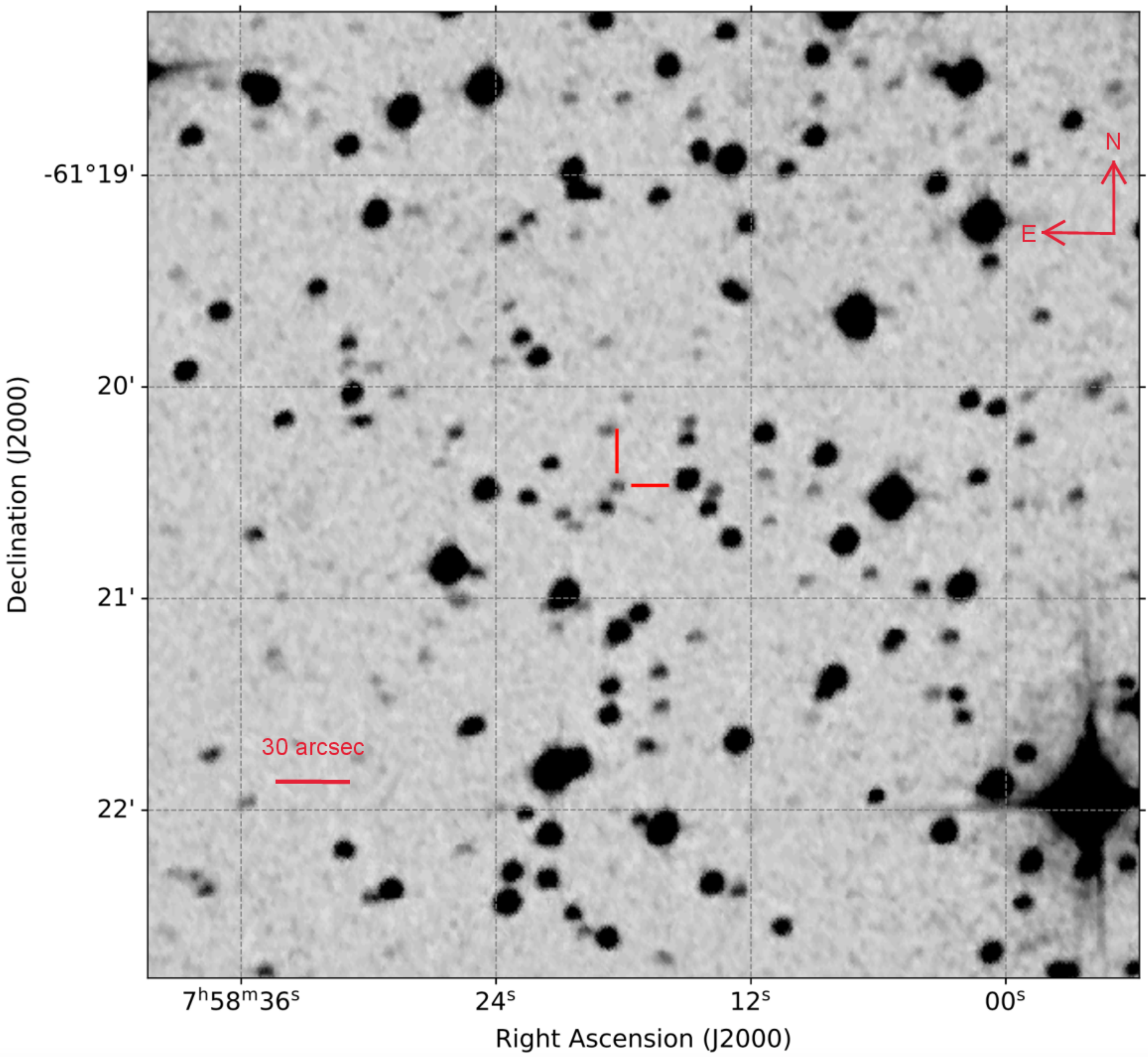}}
\caption{DSS image of the sky region around \srg. The optical counterpart is marked with red lines.
\label{f:skymap}}
\end{figure}

\section{Observations \label{s:obs}}

\subsection{Discovery with \ero}

\ero carries seven  identical cameras (TM1-TM7) each covering the same
FoV with $\sim$ 1 deg diameter.

The X-ray observations of NGC~2516 were on October 5, 2019 (OBS ID 700018, 60.6 ksec good time) and October 31, 2019 (OBS ID 700019, 80 ksec good time) in pointed mode.

Observation 700018 was carried out with only three cameras (TM5, TM6 and TM7), while 700019 was carried out with all seven cameras.

The data were reduced with the \ero Science Analysis Software System (\texttt{eSASS}) version {\tt eSASSusers\_211214 } \citep{brunner+22}.  In order to determine remaining  astrometric offsets between the two data sets, both observations were astrometrically  calibrated. Preliminary eSASS source lists were generated for both observations and corresponding {\em Gaia} counterparts were assigned within a 12 arcsec matching radius. Finding the  optimal 3-D rotation matrix $R_\mathrm{opt}$ needed to minimize the differences between X-ray positions and {\em Gaia} reference positions resembles "Wahba's problem" \citep{wahba}. Here we applied a method described in \citet{markley14} using the singular value decomposition (SVD) to find $R_\mathrm{opt}$. 

For each of the two observations the matrix $R_\mathrm{opt}$ defines an offset in the equatorial coordinates $\alpha$, $\delta$ and in the roll angle $\phi$. Neglecting the small roll angle correction, we applied the following corrections to each event:
\[ \alpha_\mathrm{corr} = \alpha_\mathrm{uncorr} + \Delta\alpha / \cos{\delta} \]
\[ \delta_\mathrm{corr} = \delta_\mathrm{uncorr} + \Delta\delta \]
with $ \Delta\alpha=-2.087$ arcsec , $\Delta\delta=+0.052$ arcsec for observation 700018 and
with $ \Delta\alpha=-2.835$ arcsec, $\Delta\delta=+4.685$ arcsec for observation 700019.

The corrected event lists were then merged and a stacked image in the energy band 0.2-5.0 keV was created (see Fig.~\ref{f:erorgb}). The new object we are reporting here, \srg, was one of the brightest X-ray sources in the field of NGC~2516 (see Fig.~\ref{f:erorgb}).

\subsection{Gaia observations}

In \textit{Gaia} DR3 \citep{gaia+21} SRGE0758 is quoted by ID 5290647986316685824 with the sky coordinates of DEC2000=119.57648825065 deg and RA2000=-61.34105844684 deg. The mean brightness of the object is 19.73$\pm0.03$, 19.90$\pm0.13$, and 18.96$\pm0.09$ in the \textit{G}, \textit{$G_{BP}$}, and \textit{$G_{RP}$} passbands, respectively. \textit{Gaia} measured the parallax of SRGE0758 as 0.648$\pm$0.35 mas. We used the geometric distance ($r_{\rm geo}$) to the system of $2733^{+4147}_{-1225}$\,pc, as determined by \cite{bailerjones+21}. The photometric distance ($r_{\rm pgeo}$) is unreliable at \textit{G}> 19 mag according to \citet{creevey+22} and \citet{fouesneau+22}. The sky position of the source is shown in Fig.~\ref{f:skymap}.

\subsection{\tes observations}

Transiting Exoplanet Survey Satellite (\tes) is providing high time-precision light curves for time-domain astrophysics \citep{ricker+14}. The satellite is equipped with four identical refractive cameras and observes the sky in sectors with 24 x 96 degrees.  Each sector is observed for two orbits of the satellite around the Earth, or about 27 days on average. Each of the 2k x 2k CCDs on the satellite has a scale of 21 arc seconds per pixel.

SRGE0758 has observations in seven sectors, that Obs. IDs of 27, 28, 31, 34, 35, 36 and 37 which are reachable from the Mikulski Archive for Space Telescopes\footnote{\protect\url{https://mast.stsci.edu}}. The observations were made between July 5, 2020, and April 02, 2021, about one year after the \ero observations and were carried out with a time resolution of 475 sec.

We extract the photometry from the TESS full-frame images using \textit{eleanor} \citep{feinstein+19}, which to obtain a cut out of 31x31 pixels postcard of the calibrated full-frame images centered on target source. \textit{eleanor} removes the background, corrects for systematic errors, and derives a light curve for different apertures.

All sectors were reduced into 11x11 pixel subgrids. We used the default pixel source extraction area that produced by \textit{eleanor} for the source to obtain time series. For each sector we plotted over Gaia DR3 measurements in the field and check the correct extraction location for the source. 

During the extraction process we encountered two revelant challenges. (1) At $G\simeq19.7$ the source is very faint, almost at background level. This indeed led in some sectors to background-subtracted light curves with negative fluxes. (2) \gai data have shown that a neighbouring source falls into the same \tes pixel. This object is $\sim$8 arcsec away from our target with coordinates RA1=119.57869730126448 and DEC1=-61.34270664883291 (source ID 5290647917596586112, $G = 19.71\pm0.02$ mag). We checked the variability of this neighbouring source from \textit{Gaia} DR3. The new \textit{Gaia} DR3 catalog released with a variability column \citep{eyer+22}. In this column, it is noted whether the system is a variable from long-term photometry. There was no indication about brightness variability of the contaminating source. 

All this means that the already weak signal from our target is further diluted by the signal from the neighbouring star, so that neither absolute brightness values nor absolute variability amplitudes can be determined. Giving those extra challenges we perform variability analysis on normalized light curves.

\begin{figure}
\resizebox{\hsize}{!}{\includegraphics{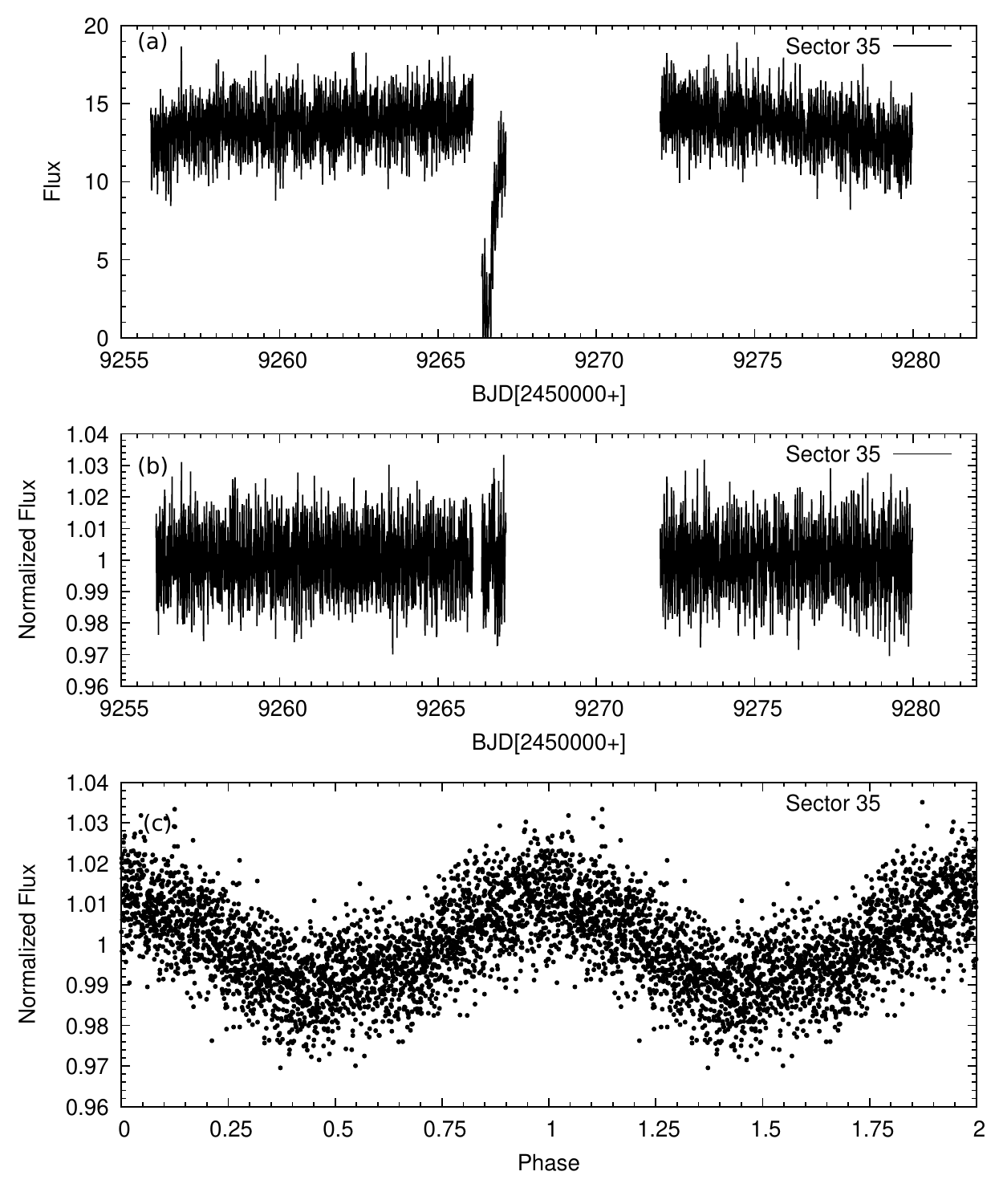}}
\caption{\tes light curve of SRGE0758 obtained for Sector 35. (a) is showing the background corrected \tes light curve in original time series. (b) shows flattened/normalised light curve for Sector 35. (c) displays folded light curve of sector 35 according to Eq.~\ref{e:eph} in normalised flux.
\label{f:tesslc}}
\end{figure}

To investigate possible periodic variations in all sectors, we applied a Savitzky–Golay smoothing filter algorithm with \textit{flatten} task in \textit{lightkurve} \citep{lcsoftware2018}. The data obtained in one of the sectors (here sector \#35) are shown in Fig.~\ref{f:tesslc} (top: original, bottom: normalized).

\begin{figure}
\resizebox{\hsize}{!}{\includegraphics{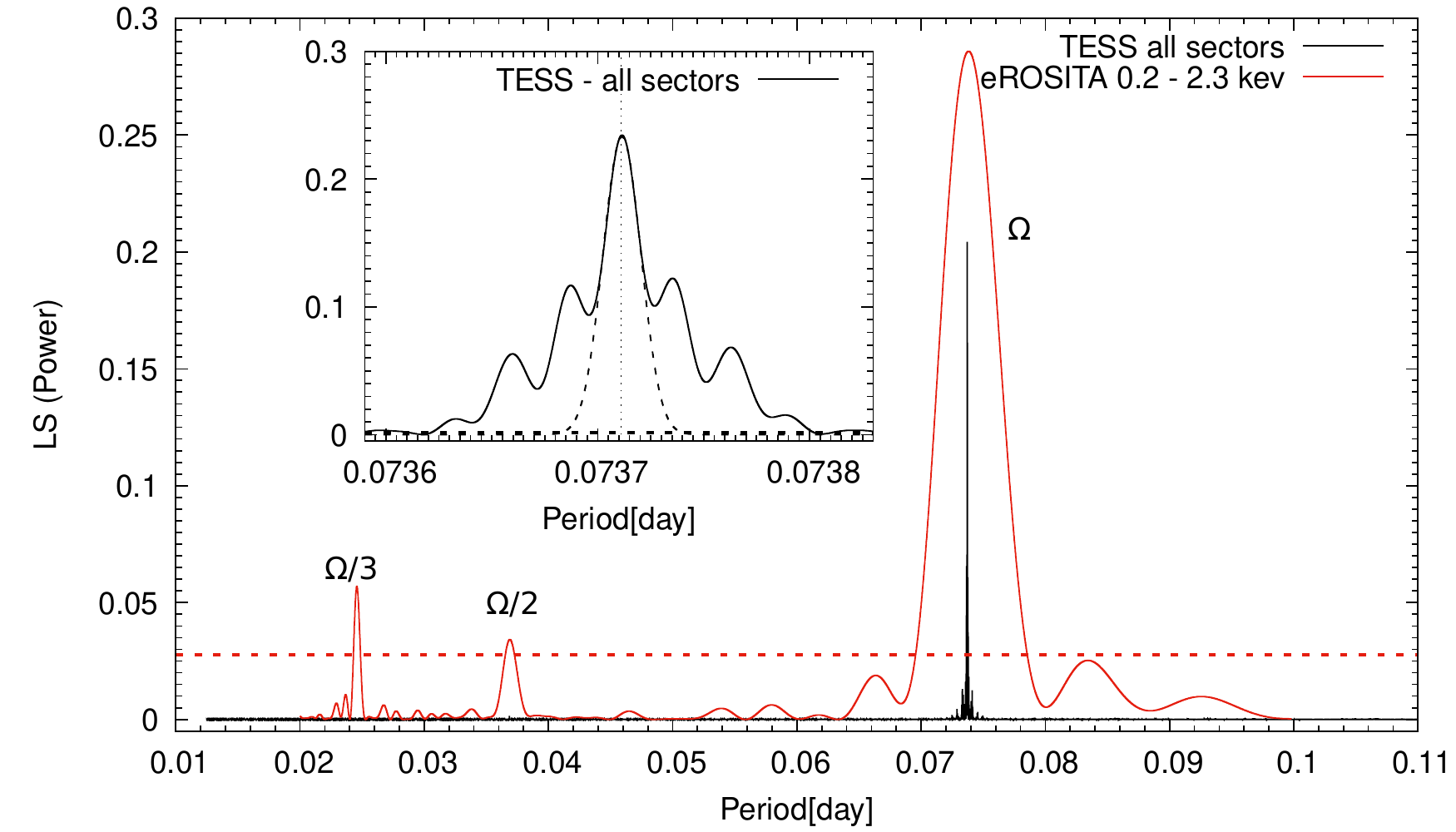}}
\caption{Lomb-Scargle periodogram of the \tes and \ero observations. Main figure shows the power spectra of the object which obtained from \tes and \ero that are over plotted together with an inset detailing the main peak of the \tes observations. Dashed black line shows a gaussian line fit to the main peak. The red dashed line in the main frame gives the 99\% significance level of the \ero power spectrum. The main period and its harmonics are indicated with the $\Omega$ symbol . The horizontal black dashed line within the inset indicates 99\% significance for the TESS spectrum.
\label{f:powspec}}
\end{figure}

\subsection{Follow-up spectroscopy with SALT}

We carried out optical spectroscopic observations of the object with the 11 meter Southern  African  Large  Telescope (SALT) at Sutherland Observatory \citep{buckley+06}. The 1200s exposure was performed with the Robert Stobie Spectrograph \citep[RSS,][]{burgh+03} on May 24, 2022. The grism 
(pg0300) covers 4000-8000 $\AA$ with a small gap at 4900-5100 $\AA$. 
The used grism together with the used slit width of 1\farcs5 resulted in a spectral resolution of 17\,\AA\ (FWHM). The spectrophotometric standard star Feige 110 \citep{oke90}, observed with SALT on 2022 May 15, was used for the flux calibration. The calibration was done in IRAF \citep{tody86} by fitting a spline3 function to the observed spectrum of the standard to determine the sensitivity function, which was then applied to the spectrum of \srg. We could only obtain relative flux calibrations, from observing a spectrophotometric standard in twilight, due to the nature of the SALT design and the moving entrance pupil \citep{buckley+18}.

\begin{figure*}
\resizebox{\hsize}{!}{\includegraphics[width=\columnwidth]{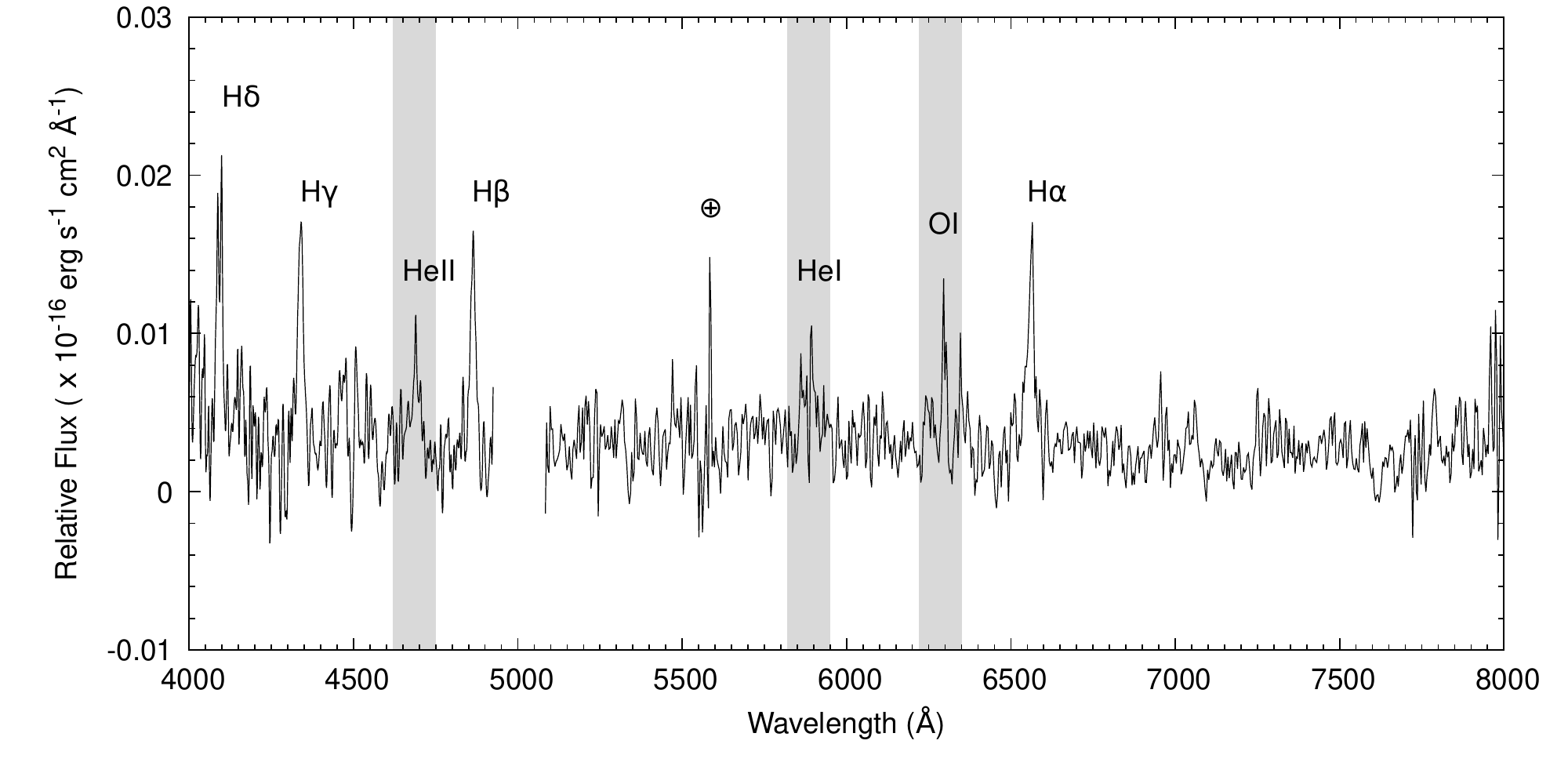}}
      \caption{The SALT spectrum of the SRGE0758. The wavelength ranges which are strongly contaminated by sky lines or cosmic rays are indicated in grey.
\label{f:saltspec}
} 
\end{figure*}

\section{Analysis and results \label{s:ana}}

\subsection{Optical photometry from \tes}

To find any periodic changes in the optical light curve, we used the Lomb \& Scargle periodogram which is useful in non-sinusoidal time series \citep[][]{lomb+76,scargle+82,vanderplas+18}. We searched for significant peaks in the frequency interval from 0 $d^{-1}$ up to the Nyquist frequency of \tes data (182 $d^{-1}$). The power spectrum shows a main peak at $P=106$\,min.
Following \cite{baluev08} we computed the false alarm probability (FAP) to check the significance of the peak. The FAP was zero, confirming that the main peak belongs to a real periodic variation. We measured the uncertainty of this period by a Gaussian fit applied to the periodogram peak. At 95\% confidence the uncertainty is $0\fd0000009$ so that the final \tes-period is $0\fd0737114(9)$. 

We applied the same method to the \ero X-ray light curve obtained in the $0.2 - 2.3$\,keV band. The resulting periodogram also shows a peak at $P=0.0738(5)$ (106.2(7)min), fully compatible with the \tes-period, so that we find  coherent periodic variations of both the optical and X-ray light curves at 106\,min. The power spectrum of the both light-curves is displayed in the Fig.~\ref{f:powspec}. No other periodicity was found in the power spectrum except for the harmonics of the main peak.

In the lower panels of Fig.~\ref{f:tesslc} we show folded \tes light curves. The original data were initially phase-folded with an arbitrary time zero. The resulting signal showed roughly a sinusoidal brightness variation in all sectors (around $\sim$5000 photometric cycles). The (arbitrary) phase of maximum brightness was determined with a Gaussian fit and the zero time corrected accordingly so that our final ephemeris for folding of data and the determination of the \tes maximum brightness reads as 

\begin{equation}
BJD(T_{0,max}) = 2459036.3484(4) + 0.0737114(9) \times E
\label{e:eph}
,\end{equation}
where the numbers in parentheses give the uncertainties in the last digits. In the following, all phases refer to the ephemeris given in Eq.~\ref{e:eph}. 

In all sectors, \tes light curves always show the same sinusoidal brightness variation but the amplitude of the sine curve changes considerably. We are working on limits of the instrument due to the relatively faint nature of the object. Therefore, the variable amplitude is due to the subtraction of different background counts/levels sector to sector. It is not possible to make an inference on the long-term variation of the object with \tes, but in short-term we obtained very strong signal that gives the same photometric period as the X-ray data in all sectors and this is quite a consistent result.

\subsection{SALT spectroscopy}

In Fig.~\ref{f:saltspec} the SALT spectrum is shown. The object displays several hydrogen Balmer emission lines superposed on a flat continuum. Other lines were tentatively detected (HeI5875, OI6300 and HeII4686) but closer inspection of the raw and sky subtracted 2D-spectra showed that these were possibly mimicked by cosmic ray hits on the CCD. We consider the information obtained from these lines unreliable and ignore them for further analysis. They are shaded grey in Fig.~\ref{f:saltspec}.

To determine the parameters of the hydrogen \textit{Balmer} lines the continuum was approximated with a low-order polynomial and the spectrum normalized to this continuum. Line parameters were determined by single Gaussians fits to the data. They are listed in Table~\ref{t:1}. The signal-noise ratio is low at the blue end of the spectrum and the measurement of the H$\delta$ line in particular rather uncertain. It has a double-peaked shape, different from the other lines, which is regarded being apparent only. Its measurements are regarded controversial and indicated by a colon in Table~\ref{t:1}.

From flat continuum at 6250\,\AA\, the mean spectral flux density is about 6.5$\times$10$^{-18}$ erg cm$^{-2}$ s$^{-1}$ \mbox{\,\AA}$^{-1}$, which corresponds to a \textit{Gaia} G-band magnitude of 21.5 (Vega). Hence, at the time when the spectrum was obtained the source appeared considerably fainter than found previously with \gai, $G \sim 19.7$ mag). The spectral exposure was started at JD 2459724.2284838, the spectrum thus covered the phase interval $0.07 - 0.26$, the decreasing branch of the \tes light curve.

Strong emission lines of hydrogen and helium, both neutral and ionized, are main indicators to identify a source as a CV \citep{szkody98}. We have discovered with certainly only hydrogen Balmer emission lines. This and the low brightness at the time of the SALT observation is indicative of a low accretion state of the object during our observations. Low accretion states are regularly found in strongly magnetic CVs, the polars. In these states typically only hydrogen Balmer lines (in particular H$\alpha$) are visible \citep{schwope+02}. The SALT spectroscopy thus suggests the identification of \srg as a polar in a low accretion state.

\subsection{eROSITA observations}

\subsubsection{X-ray photometry \label{s:xobs}}

\begin{figure*}
\resizebox{\hsize}{!}{\includegraphics[width=\columnwidth]{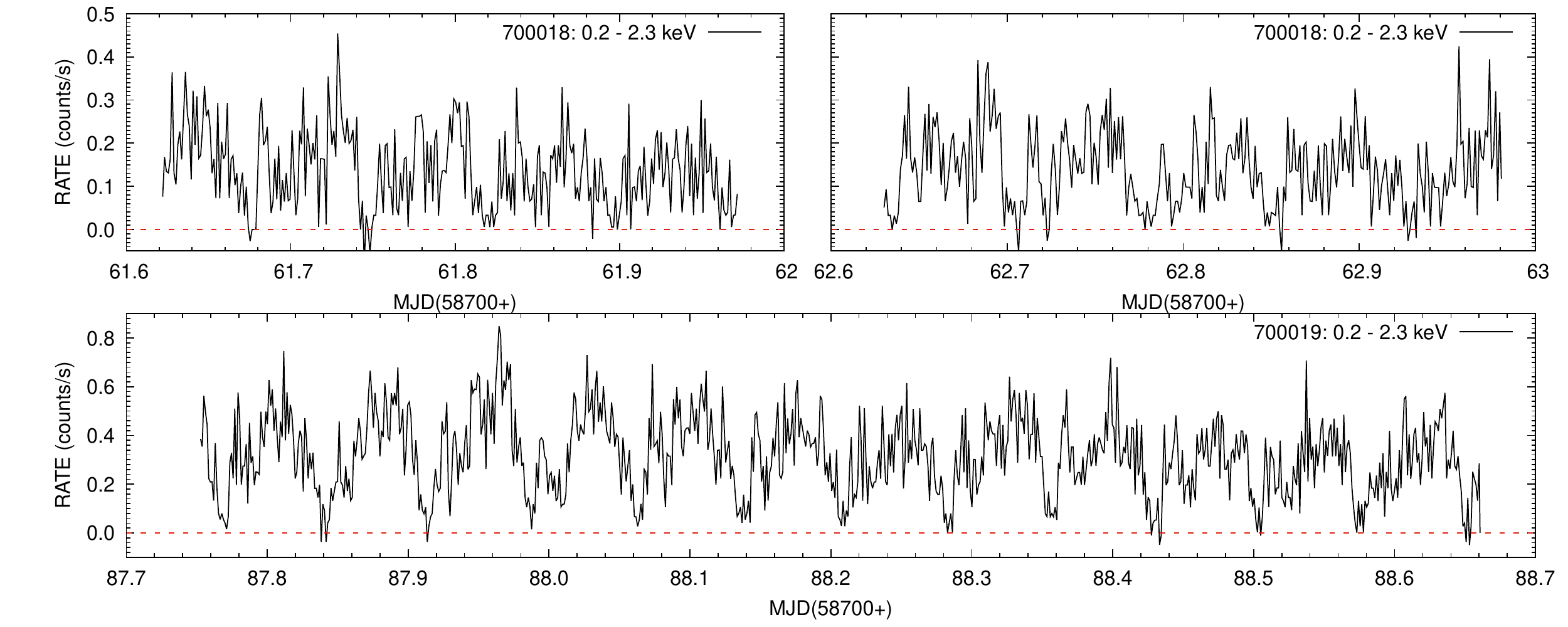}}
\caption{\ero $0.2-2.3$\,keV X-ray light curve of \srg with time bins of 100\,s. The top row shows the data from ObsID 700018 the lower row that of ObsID 700019.
\label{f:erolc}}
\end{figure*}

\ero X-ray light curves of both observations of SRGE0758 are shown in original time sequence in Fig.~\ref{f:erolc}. Only data in the interval $0.2 - 2.3$ keV and time bins of 100\,s were used. The object showed pronounced brightness variations with one bright hump and two dip or eclipse-like brightness minima. While the main minimum can always clearly recognized, the second one following in time is quite shallow. This overall variability pattern is visible during all 22 cycles that were covered during the X-ray observations. The main minimum has sometimes zero count rate but remains finite at other instancies. Hence, it is not interpreted as a true eclipse, were the count rate would always be consistent with zero.

\begin{figure}
\resizebox{8cm}{!}{\includegraphics[width=\columnwidth]{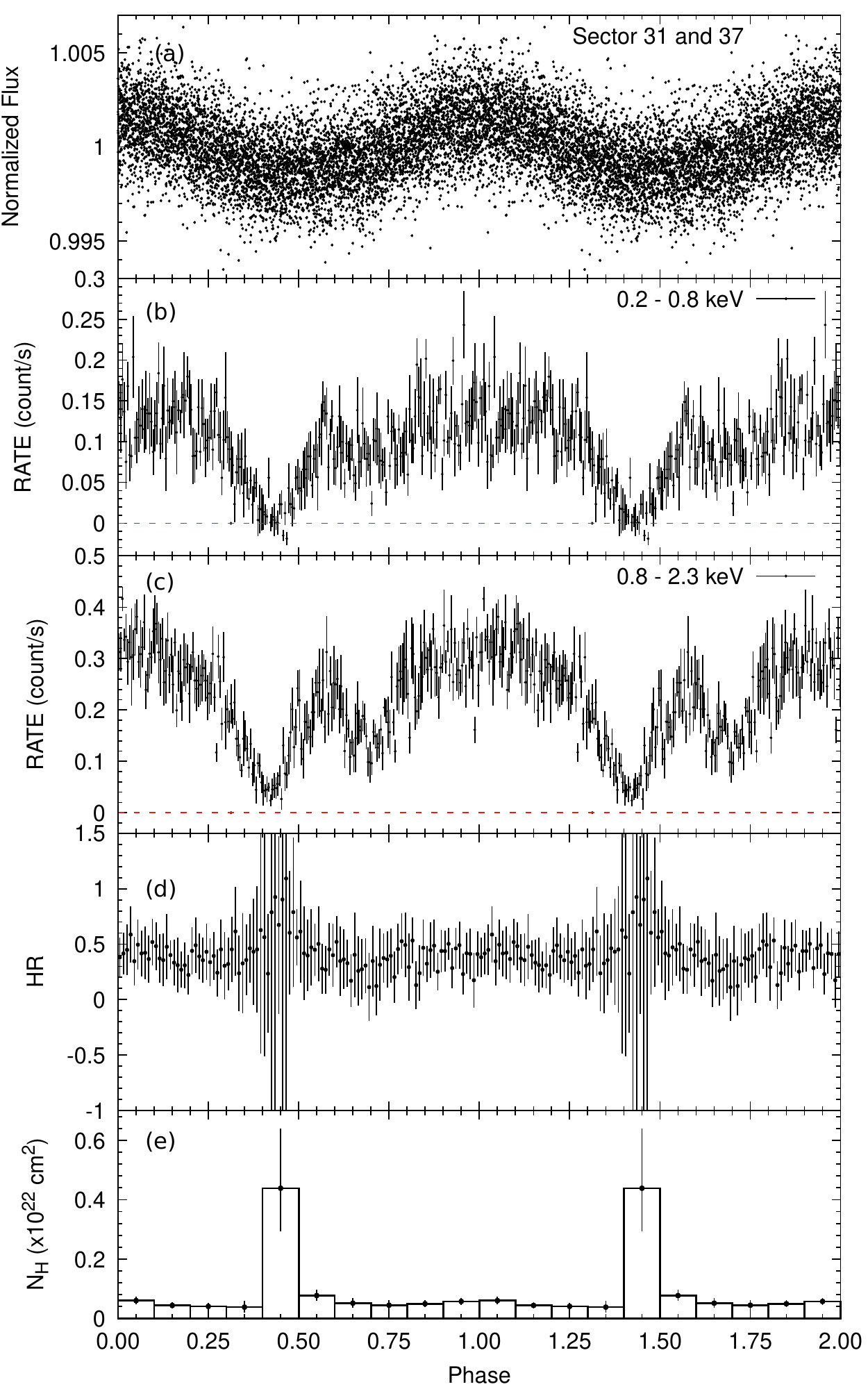}}
      \caption{\tes, energy resolved \ero light curves, \textit{Hardness Ratio} and $N_{\rm H}$ variations of SRGE0758 for ObsID 700019. (a) \tes light curve obtained from sectors 31 and 37. (b) and (c): the X-ray data in the two indicated energy bands were originally binned using 50 sec time bins,  and eventually averaged using a phase bin size of 0.005 phase units. (d) hardness ratio obtained from 0.2 - 0.8 keV and 0.8 - 2.3 keV energy bands. (e) column density variation. All light curves folded according to photometric period and T$_{0,max}$ in Eq.~\ref{e:eph}.
\label{f:hr}
}
\end{figure}

We generated energy-resolved and phase-folded X-ray light curves in two energy bands, a soft band between 0.2 - 0.8 keV and a hard band between 0.8 - 2.3 keV. These are shown in Fig.~\ref{f:hr}(b) and Fig.~\ref{f:hr}(c). The first dip is located at phase 0.40 and has a FWHM of 0.07 phase units. The flux is almost zero in the soft band ($0.2 - 0.8$\,keV), but stays always positive in the hard band ($0.8 - 2.3$\,keV). The second dip shows a contrasting behaviour. The dip is centered on phase 0.69, i.e. 0.29 phase units after the first one. It is shallow in soft X-rays and more pronounced and deeper in the hard band.  
\begin{table}
\centering
\caption{Spectral parameters of Balmer emission lines in the SALT spectrum. Flux units are 10$^{-16}$ erg cm$^{-2}$ s$^{-1}$. A colon indicates an uncertain measurement.}
\label{t:1}
\resizebox{\columnwidth}{!}{
\begin{tabular}{llcccc}
Lines &  Observed ($\AA$)   &Shift ($\AA$) & FWHM ($\AA$)  &   EW($\AA$) & Flux \\
\hline
\hline 
H$\alpha(6563\lambda)$   & 6560.7$\pm1.2$&  3.0    & 23.7  &  -66.6 & 3.7   \\
H$\beta (4861\lambda)$   & 4863.4$\pm0.9$&  2.4  & 14.7  &  -42.2 & 2.5   \\
H$\gamma(4340\lambda)$   & 4338.5$\pm1.0$& -1.5  & 19.3  &  -63.3 & 3.4   \\
H$\delta(4101\lambda)$   & 4092.9$\pm1.8$:&-8.1: & 20.1:  &  -55.4: & 3.7:   \\
\hline 
\hline 
\end{tabular}
}
\end{table}

To gain a better understanding of these variations, we calculated the Hardness Ratio (\textit{HR}) between the two chosen bands. HR is defined as $HR = (H-S)/(H+S)$ and thus can vary between $-1$ and $+1$ for supersoft and superhard sources (see Fig.~\ref{f:hr}(d)). Outside the dips the HR shows little variability and scatters around $\sim$0.5. 


The dips (or eclipse-like features) could be related to photoelectric absorption or to geometric effects like self-eclipses and/or geometric foreshortening of the corresponding emission regions.

The observed spectral hardening during the first dip implies that it is likely caused by photo-electric absorption. The absorber is the magnetically guided stream or curtain out of the orbital plane like those studied by e.g.~\cite{watson+89} and also described by \cite{kuulkers+10}. The energy-independent variation during the second dip suggests that it is more likely caused by foreshortening/self-eclipse of the accretion/emission region as the white dwarf rotates. The self-eclipse is not total but partial only, which can be explained if the emission region is extended and not completely vanishing behind the limb of the WD as it rotates. 

\subsubsection{X-ray spectroscopy \label{s:hr}}

Spectral analysis was performed with the XSPEC package \citep[version 12.12.0][]{arnaud+96,dorman+01}. Only photons between $0.2 - 5$ keV were included in the spectral analysis. Higher energies were completely dominated by noise. We started out by selecting photons from the bright phase between $\phi = 0.8 - 1.2$ only  (see Fig~\ref{f:hr}). Photons were grouped with a minimum of 20 counts per bin and we utilised \textit{chi} statistics for optimizing the fit.

Initially we used just a single thermal emission model \citep{mewe+85,liedahl+95} with solar abundances \citep{wilms+00} absorbed by cold interstellar matter of some column density $N_{\rm H}$, so that our spectral model reads as {\tt TBABS * MEKAL} in XSPEC terms. The object is relatively distant, the inverse parallax is $\sim$1500\,pc while the estimate by \cite{bailer-jones+21} points to an even larger distance of $\sim$2733 pc. Accordingly, we may expect that the X-ray spectrum 'sees' a large fraction of the whole galactic column in this direction which is $N_{\rm H,gal}$ = $1.26\times10^{21}$\,cm$^{-2}$.

\begin{figure}
   \includegraphics[width=\columnwidth]{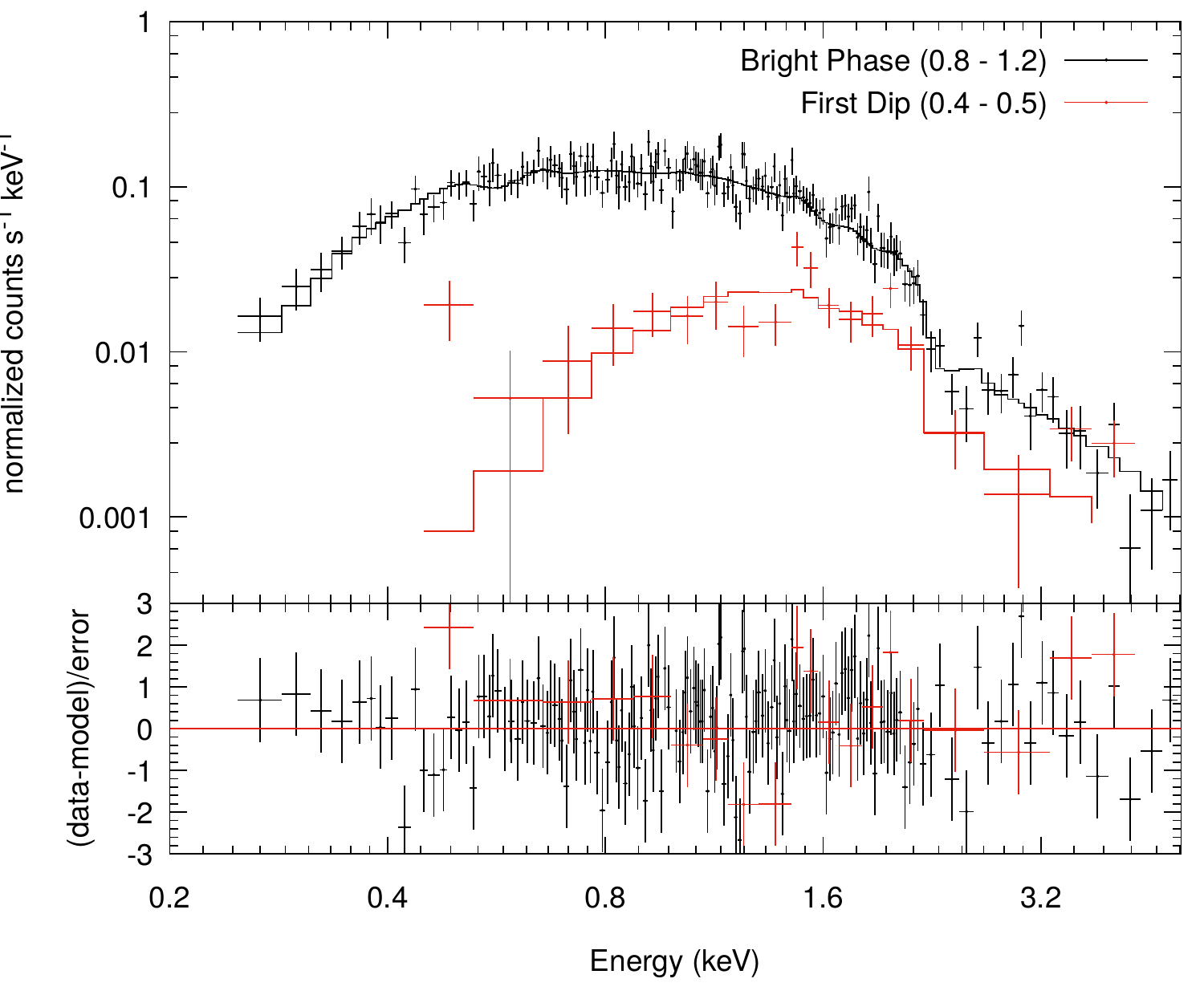}
      \caption{Phase resolved X-ray spectra of the SRGE0758. The black data points represent the bright phase spectrum and were extracted for the phase interval $0.8 - 1.2$. The black line shows the model fit with a reduced chi-square value of 1.02 ($\chi = 171$ for 168 dof). The spectrum of the first dip (shown with red data points) was obtained from phase at $0.4 - 0.5$, the fit reveled a reduced $\chi^2$ of 1.51 ($\chi = 30$ for 20 dof).} 
      \label{f:brightpha}
\end{figure}

The simple model already gave a good fit to the data with $\chi^2 = 171$ for 168 degree of freedom, $\chi^2_\nu = 1.02$ (see Fig.~\ref{f:brightpha}).

Many \ros-discovered polars show a soft, blackbody-like, radiation component from the bottom of the accretion column and its surroundings. In contrast, we could not find evidence of such a blackbody component in the spectrum.

We calculated the unabsorbed fluxes with the \textit{cflux} task in XSPEC. After ideal spectral fitting, we calculated the bolometric X-ray flux with XSPEC using the best-fit spectral parameters and a dummy response which covers the broad energy range between 10$^{-6}$ and $10^4$\,keV. The bolometric correction factor is 1.28. Errors were calculated with the \textit{error} command in XSPEC (90\% significance). All the fit results and uncertainties are listed in Table~\ref{t:2}. 


We then wanted to better constrain the spectral parameters as a function of the spin-phase, in particular the change in the column density through the main absorption dip. We therefore generated phase-resolved spectra using 10 phase bins of equal length. The photons per phase bin were grouped again with 20 counts per spectral bin. The temperature of the {\tt MEKAL} model was fixed at the bright-phase temperature so that only the column density and the model normalization were allowed to vary. In the bottom panel of Fig.~\ref{f:hr} the phase-dependent behaviour of the column density is shown. It was $N_{\rm H}=0.5\times10^{21}$\,cm$^{-2}$ for most of the phase bins as for the mean spectrum. During the first dip a marked increase by almost a factor 10 (although with large error bar) was observed which was even in excess of the galactic column density. This accretion-related high column density is very similar or comparable with other well known polar, HU Aqr during its high accretion state (factor of 10) \citep{schwarz+09}.

We may derive a rough estimate of the mass accretion rate limit via the X-ray luminosity. If one assumes that the WD in \srg is a typical one, its likely mass is 0.8 $M_{\odot}$ \citep{pala+21}. We estimated $R_{\rm wd}$ for this mass using the \citet{nauenberg+72} mass--radius relation. If one further assumes that the X-ray luminosity is a good proxy for the accretion luminosity, $L_{\rm acc} \simeq L_{\rm X}$, one gets $\dot{M}=L_{\rm X}R_{\rm 0.8}/M_{\rm 0.8}G = 8.3(7)\times10^{-11}$ $M_{\odot}$ yr$^{-1}$ (only flux error considered here). The accretion rate was calculated for the distance of 2733 pc \citep{bailerjones+21}. These authors also give a rather large confidence interval between 1225 pc and 4147 pc. Taking the uncertainty in the distance into account, the likely accretion rate lies in the interval $\dot{M} \approx 2 - 19 \times10^{-11}$ $M_{\odot}$ yr$^{-1}$

\subsubsection{\ero all sky surveys \label{s:erass}}

In Fig.~\ref{f:erass14} X-ray images for the first four \ero All-Sky Survey (eRASS) scans of \srg are shown. The new object was detected in the first three scans but not in the last (eRASS4). The mean count rates and observation times are listed in Table~\ref{t:3}. Because of the high ecliptic latitude of the object, every survey scans the position of \srg quite often every four hours thus building up a farther deep survey exposure \citep{freyberg+20}. 

We calculate the count rate upper limit at the position of the source in eRASS4 based on X-ray aperture photometry following the Bayesian approach described by \cite{1991ApJ...374..344K}. We collect observed counts, background counts, and exposure time within a circular aperture with a radius of $r_{\rm ape}=30.5$ arcseconds at the 0.2 - 2.3 keV energy band. The aperture corresponds with a PSF encircled energy fraction of $\rm EEF=0.75$, which measures the fraction of total photons within the aperture.

The X-ray flux decrease in eRASS1 and non-detection in eRASS4 suggest overall X-ray emission cut in the object which indicate different mass accretion state.

\bgroup
\def\arraystretch{1.10}
\begin{table}
\centering
\caption{Spectral parameters from bright phase interval ($\phi$=0.8 - 1.2): spectral fit parameters, their uncertainties, fit statistics, and model bolometric fluxes.}
\label{t:2}
\resizebox{\columnwidth}{!}{
\begin{tabular}{lcc}
\hline       
\hline                     
\multicolumn{2}{l} {Model: {\tt TBABS*(MEKAL)}}           \\
\hline
Parameters               &                                \\
\hline
   
$N_{\rm H}$($10^{22}$cm$^{2}$)  & $0.05^{+0.01}_{-0.01}$   \\
$kT_{\rm mekal}$(keV)            & $9.7^{+7.7}_{-2.6}$     \\

$\chi^{2}$(d.o.f)            & 1.02(171/168)             \\
\hline
Unabsorbed Fluxes $(10^{-13}\fergs)$                   \\
\hline
$F_{0.5-2.5} $  & $3.94^{+0.02}_{-0.02}$   \\
$F_{2.5-10}  $  & $7.2^{+0.8}_{-1}$        \\
$F_{\rm bol}     $  & $17.7^{+1.6}_{-1.6}$ \\
\hline
$L_{x}$(erg s$^{-1}$)$(10^{32})$ & $7.9(7)$      \\
\hline
\end{tabular}
}
\end{table} 
\egroup


\subsection{Accretion geometry}

The X-ray light curve of the object does not contains an eclipse of a WD. Therefore we do not know the true orbital phase zero. The X-ray observations only reveal one bright hump with two consecutive dips. Conformably, optical variability obtained with \tes revealed the same photometric period. The optical light curve shows just one hump. Under these circumstances, it is not possible to specify the location of the accretion column precisely. But we can nevertheless derive a few constraints.

\begin{figure}
   \includegraphics[width=\columnwidth]{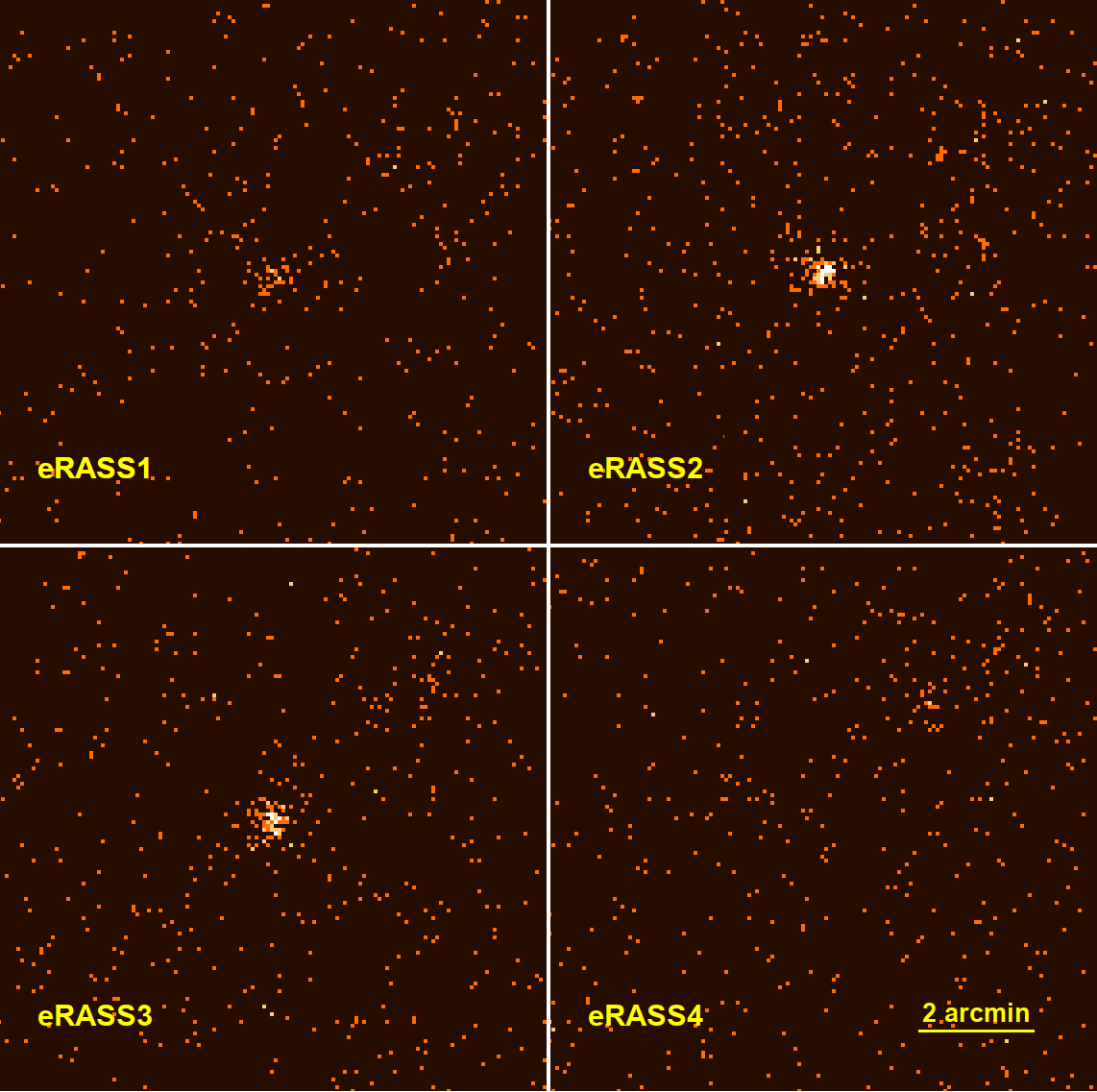}
      \caption{Images of the CV during the 4 \ero All Sky Surveys in the 0.2–2.3 keV range. During eRASS4,the object is in a low state.} 
      \label{f:erass14}
\end{figure}

\begin{table}
\centering
\caption{Observation log of the \ero All Sky Surveys and pointed observations covering the location of \srg. The given times (MJD) refer to the middle of the concerned observation. }
\label{t:3}
\resizebox{\columnwidth}{!}{
\begin{tabular}{lllcc}
Survey &  Count Rates (s$^{-1}$)   & Time (MJD) & Total Scans & Total Exposure (s)\\
\hline
\hline 
eRASS1       & 0.11$\pm0.02$& 59002.265 &22 & 634     \\
eRASS2       & 0.27$\pm0.03$& 59188.520 &24 & 736     \\
eRASS3       & 0.27$\pm0.03$& 59371.505 &26 & 810     \\
eRASS4       & $< 0.038$ (99.87\%)&        & 25 & 764 \\
700018(CalPV)& 0.244$\pm0.005$& 58762.302 & - & 60.6k \\
700019(CalPV)& 0.277$\pm0.006$& 58788.207 & - & 80k   \\

\hline 
\hline 
\end{tabular}
}
\end{table} 


The increased absorption at the phase of the deep dip suggests an origin due to photo-electric absorption in the accretion stream. The second dip is likely caused by a self-eclipse of the accretion column by WD itself. During the second dip the accretion region is partially hidden behind the limb of the WD. Accordingly, this accretion region which is visible for most of the orbital cycle is located in the upper hemisphere (with respect to the orbital plane, $\beta$<90$^{\circ}$ with $\beta$ being the co-latitude of the accretion region). The object does not show an eclipse, hence the inclination is constrained to $i<75\degr$. 

The stream eclipse at 0.4 phase can only occur if the orbital inclination is larger than the co-latitude of the column ($i>\beta$). If our interpretation of the second dip as due to a self-eclipse is correct, then $i + \beta \simeq 90\degr$.
The mid points of the two dips are separated by $\Delta\phi\sim$0.29 phase. Best visibility of the accretion region is then expected at half a cycle later, i.e.~about 0.2 phase units before the first dip. One would expect then to observe a clear maximum in the light curve due to the given best visibility of the accretion region which is not clearly observed. Instead there is an extended bright phase without a clear center or maximum.

The typical location of a stream dip in polars is between phases 0.8 and 0.9 \citep[e.g.]{schwope+01}. In polars with a pronounced bright-phase hump in the light curve the dip is often located in the second half of the bright phase, i.e.~it occurs at bright phase center of shortly thereafter. If our tentative geometry is correct, the situation in \srg is rather different with the dip trailing the bright-phase by 0.2 phase units. 
Such a geometry seems to be unusual but actually there are other polars which display similar photometric features, namely QQ Vul \citep{belle+00} and V834 Cen \citep{schwope+08}. The true orbital phase zero of these systems are known. In both objects true phase zero was located between the two consecutive dips. 

\begin{figure}
   \includegraphics[width=\columnwidth]{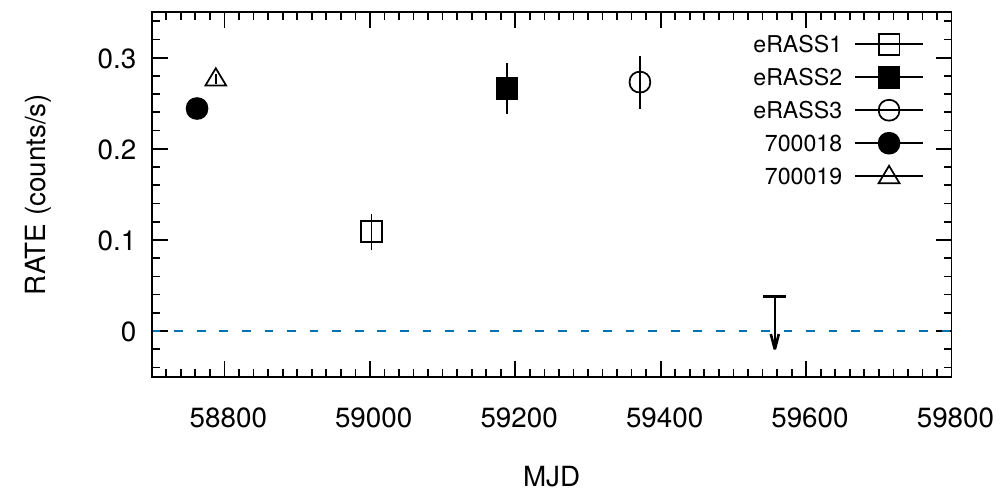}
      \caption{Long-term X-ray light curve of SRGE0758 which corresponds to eRASS1-4 and pointed \ero observations. Black arrow displays the upper flux limit of the eRASS4 survey for given sky position of the SRGE0758.} 
      \label{f:erass14}
\end{figure}

\section{Discussion and conclusion \label{s:dis}}

We have analysed extended \ero and \tes observations of a serendipitously discovered X-ray source in the direction of the open cluster NGC~2516, that clearly does not belong to the cluster. It is among the X-ray brightest objects in the \ero CalPV observation of the field but is an accretion-powered background object at a distance of about 1225 to 4147 pc. The cluster itself is at a distance of 400 pc \citep{terndrup+02}. We also obtained SALT spectroscopic follow-up and time-resolved photometry from the SAAO. \ero X-ray and optical \tes observations reveal a periodic variability of the source at 106 min. This is the only found period in the system. The identification spectrum was obtained when the object was $\sim$2 mag fainter compared to its \gai measurements. The spectrum reveals hydrogen Balmer emission superposed on a flat or slightly blue continuum. All these features clearly identify the new object as a magnetic cataclysmic variable. It likely belongs to the AM Herculis or polar subclass although a nature as an \textit{Intermediate Polars} (IP) could not ruled out completely. This identification is based on the occurrence of just one period in the system, the spin period of the white dwarf, and the accretion rate changes which are typically observed in polars. Also, the low state spectrum is similar to other polars. 

The X-ray light curves are modulated by 100\% on the 106 min period. The most structuer in the light is an eclipse-like broad dip at phase 0.40 which we interpret as stream absorption. Stream eclipses are common among the polars, they are e.g.~observed in EF Eri \citep{beuermann+91}, EK UMa \citep{clayton+94}, QQ Vul \citep{beardmore+95}, HU Aqr \citep{schwope+01} and UZ For \citep{warren+95}, and require that the magnetically guided accretion stream lifted out of the orbital plane crosses the line of sight. The second dip is clearly revealed in the phase-folded X-ray light curve, not so much in the individual cycles, and is tentatively identified as a possible self-eclipse of the emission region, because no change of the hardness ratio or the column density was observed at this phase. The self-eclipse could be grazing only, the X-ray flux stays finite at this phase. This interpretation is based on the observation that neither the hardness ratio nor the column density in the spectral fits show significant variability at this phase. 

Due to the lack of a primary star eclipse or other information from e.g. spectral line variability the true phase zero is unknown. But the fact that the 106-minute photometric period is the same in both \tes and \ero observations suggesting a synchronous spin/orbital rotation, a typical property of polars \citep{warner+95}. The large portion of the polar population has orbital periods less than 2 hours \citep{kuulkers+10} as our new object, and also we did not encounter any change implying the white dwarf spin seen as in \textit{Intermediate polars}. In IPs, the accretion can be occur in two way, stream-fed or disc-fed. In both accretion scenario, the matter changes poles due to the WD spin and this breeds as itself a beat period in periodograms obtained from X-ray and optical light curves \citep{norton+97}. 

The mean optical light curve from \tes is not as complex as the X-ray light curve. The variation is very smooth and almost sinusoidal. The variability amplitude is likely small but cannot properly quantified given the relative faintness of the star and the contamination due to a second object in the same pixel. The maxima of the \tes and \ero light curves are roughly coincident. We could not detect any sharp and strong accretion stream or curtain related absorption in the \tes light curves. The \tes minimum lies between the two dips in X-rays but closer to the first. Optical variability in polars may be caused by cyclotron beaming, geometric foreshortening of a heated accretion region and obscuration of the light sources. Given the lack of detailed light curve information, we do not speculate about the origin of \tes variability in \srg.

The X-ray spectrum is compatible with single temperature thermal plasma emission at around 10 keV, well in line what is typically found in polars \citep{kuulkers+10}. Polars have lower X-ray temperatures than IPs \citep{mukai17}. The new object lacks a soft black body component in its X-ray spectrum which was often prominent in the X-ray spectra of \ros-discovered polars \cite{beuermann+schwope94}. This is similar to other polars discovered with \xmmn \citep{vogel+08,ramsay+09,webb+18} and also the first \ero-discovered polar seem to lack this component \cite{schwope+22}. 
The strength of the soft X-ray excess in polars and has been a hotly debated topic \citep[e.g.][]{vogel+08,ramsay+09}. It seems that the effective energy range of \xmmn, which cannot extend to EUV wavelengths, makes it difficult to detect this component of re-processed emission from the WD in some systems. Presumably, we are facing with the same issue in \ero which has a similar sensitivity range.


The optical identification spectrum has a flat continuum and contains broad hydrogen emission lines which is a good indicator for magnetic accretors. The continuum neither reveals the white dwarf not cyclotron harmonic emission nor the donor star. A detection of the donor cannot be expected given large distance and the short orbital period, hence late spectral type \citep[][predict a spectral type later than M5.3]{knigge+11} and thus faintness of the star. Another specific feature needs to be argued here is the large scale brightness change between \gai and SALT spectrum ($\sim$1.8 mag). While in polars orbital phase dependent variability of up to 4 magnitudes may occur, the brightness drop by about 2 mag between \gai and SALT is more likely due a reduced mass accretion rate. The lack of clearly detected helium lines is suggestive of a reduced accretion state. The non-detection in eRASS4, about 5 month prior to the SALT observations support this view. Accretion state transitions occur randomly in polars. The time they spend in these states is also randomly changing and unpredictable. The triggering mechanism is thought to be predominantly chromospheric activity in the donor star \citep{kafka+05}. On the other hand, accretion state variations are not a unique indication of a polar. Also the intermediate polars are meanwhile frequently being discovered in low states, the first perhaps being FO Aqr \citep[e.g.][]{littlefield+20,kennedy+20} followed by several others\citep[e.g.][]{covington+22,hill+22}. 


The position of the first dip with respect to the main X-ray hump in the light curve implies that the object has a very interesting accretion geometry. The column is not in the direction of the secondary star. The appearance of the stream dip after the main hump implies that the accretion spot is in a rather perpendicular position to the system. The accretion to the distant/main pole is more likely occur at high accretion states \citep{ferrario+89} in polars.


Previously, \textit{ROSAT} and \textit{Swift} observatories observed the NGC~2516 but \srg could not be detected by either mission. 
Probably, the object was in a low accretion state during these observations.
The sensitivity of \ero together with its optimized scanning strategy led to the discovery of this object despite its large distance and the switch between high and low states. It may be expected the \ero will uncover many more magnetic CVs through systematic follow-up observations of all point-like X-rays sources found in the eRASS. In  particular there is a great chance to find those that escaped its detection in the \ros all-sky survey while being in a low state at that time, because the duty of polars might be as short as 50\% \citep{hessman+00}.

\begin{acknowledgements}
Samet Ok is supported by TUBITAK 2219-International Postdoctoral Research Fellowship Program for Turkish Citizens. Samet Ok acknowledge deep gratitude to the Leibniz Institute for Astrophysics Potsdam (AIP) and members of the X-ray astronomy group for the warm hospitality during his stay.
This work was supported by the Deutsches Zentrum f\"ur Luft- und Raumfahrt (DLR) under grants 50 QR 2104, 50 OX 1901, and 50 OR 2203. Support by the Deutsche Forschungsgemeinschaft under grant Schw536/37-2 is gratefully acknowledged.
This research has made use of data, software and/or web tools obtained from the High Energy Astrophysics Science Archive Research Center (HEASARC), a service of the Astrophysics Science Division at NASA/GSFC and of the Smithsonian Astrophysical Observatory’s High Energy Astrophysics Division. 
This paper includes data collected by the \tes mission. Funding for the \tes mission is provided by the NASA's Science Mission Directorate. This work is based on data from eROSITA, the soft X-ray instrument aboard SRG, a joint Russian-German science mission supported by the Russian Space Agency (Roskosmos), in the interests of the Russian Academy of Sciences represented by its Space Research Institute (IKI), and the Deutsches Zentrum für Luft- und Raumfahrt (DLR). The SRG spacecraft was built by Lavochkin Association (NPOL) and its subcontractors, and is operated by NPOL with support from the Max Planck Institute for Extraterrestrial Physics (MPE). The development and construction of the eROSITA X-ray instrument was led by MPE, with contributions from the Dr. Karl Remeis Observatory Bamberg \& ECAP (FAU Erlangen-Nuernberg), the University of Hamburg Observatory, the Leibniz Institute for Astrophysics Potsdam (AIP), and the Institute for Astronomy and Astrophysics of the University of Tübingen, with the support of DLR and the Max Planck Society. The Argelander Institute for Astronomy of the University of Bonn and the Ludwig Maximilians Universität Munich also participated in the science preparation for eROSITA. The eROSITA data shown here were processed using the eSASS software system developed by the German eROSITA consortium. We thank the anonymous referee for useful comments
and suggestions which helped to improve the paper.

\end{acknowledgements}

%
%

\bibliography{srgej075818}

\begin{thebibliography}{69}
\expandafter\ifx\csname natexlab\endcsname\relax\def\natexlab#1{#1}\fi

\bibitem[{{Arnaud}(1996)}]{arnaud+96}
{Arnaud}, K.~A. 1996, in Astronomical Society of the Pacific Conference Series,
  Vol. 101, Astronomical Data Analysis Software and Systems V, ed. G.~H.
  {Jacoby} \& J.~{Barnes}, 17

\bibitem[{{Bailer-Jones} {et~al.}(2021{\natexlab{a}}){Bailer-Jones}, {Rybizki},
  {Fouesneau}, {Demleitner}, \& {Andrae}}]{bailer-jones+21}
{Bailer-Jones}, C.~A.~L., {Rybizki}, J., {Fouesneau}, M., {Demleitner}, M., \&
  {Andrae}, R. 2021{\natexlab{a}}, \aj, 161, 147

\bibitem[{{Bailer-Jones} {et~al.}(2021{\natexlab{b}}){Bailer-Jones}, {Rybizki},
  {Fouesneau}, {Demleitner}, \& {Andrae}}]{bailerjones+21}
{Bailer-Jones}, C.~A.~L., {Rybizki}, J., {Fouesneau}, M., {Demleitner}, M., \&
  {Andrae}, R. 2021{\natexlab{b}}, VizieR Online Data Catalog, I/352

\bibitem[{{Baluev}(2008)}]{baluev08}
{Baluev}, R.~V. 2008, \mnras, 385, 1279

\bibitem[{{Beardmore} {et~al.}(1995){Beardmore}, {Ramsay}, {Osborne}, {Mason},
  {Nousek}, \& {Baluta}}]{beardmore+95}
{Beardmore}, A.~P., {Ramsay}, G., {Osborne}, J.~P., {et~al.} 1995, \mnras, 273,
  742

\bibitem[{{Belle} {et~al.}(2000){Belle}, {Howell}, \& {Mills}}]{belle+00}
{Belle}, K.~E., {Howell}, S.~B., \& {Mills}, A. 2000, \pasp, 112, 343

\bibitem[{{Beuermann} \& {Schwope}(1994)}]{beuermann+schwope94}
{Beuermann}, K. \& {Schwope}, A.~D. 1994, in Astronomical Society of the
  Pacific Conference Series, Vol.~56, Interacting Binary Stars, ed. A.~W.
  {Shafter}, 119

\bibitem[{{Beuermann} {et~al.}(1991){Beuermann}, {Thomas}, \&
  {Pietsch}}]{beuermann+91}
{Beuermann}, K., {Thomas}, H.~C., \& {Pietsch}, W. 1991, \aap, 246, L36

\bibitem[{{Brunner} {et~al.}(2022){Brunner}, {Liu}, {Lamer}, {Georgakakis},
  {Merloni}, {Brusa}, {Bulbul}, {Dennerl}, {Friedrich}, {Liu}, {Maitra},
  {Nandra}, {Ramos-Ceja}, {Sanders}, {Stewart}, {Boller}, {Buchner}, {Clerc},
  {Comparat}, {Dwelly}, {Eckert}, {Finoguenov}, {Freyberg}, {Ghirardini},
  {Gueguen}, {Haberl}, {Kreykenbohm}, {Krumpe}, {Osterhage}, {Pacaud},
  {Predehl}, {Reiprich}, {Robrade}, {Salvato}, {Santangelo}, {Schrabback},
  {Schwope}, \& {Wilms}}]{brunner+22}
{Brunner}, H., {Liu}, T., {Lamer}, G., {et~al.} 2022, \aap, 661, A1

\bibitem[{{Buckley} {et~al.}(2018){Buckley}, {Andreoni}, {Barway}, {Cooke},
  {Crawford}, {Gorbovskoy}, {Gromadzki}, {Lipunov}, {Mao}, {Potter},
  {Pretorius}, {Pritchard}, {Romero-Colmenero}, {Shara}, {V{\"a}is{\"a}nen}, \&
  {Williams}}]{buckley+18}
{Buckley}, D. A.~H., {Andreoni}, I., {Barway}, S., {et~al.} 2018, \mnras, 474,
  L71

\bibitem[{{Buckley} {et~al.}(2006){Buckley}, {Swart}, \&
  {Meiring}}]{buckley+06}
{Buckley}, D. A.~H., {Swart}, G.~P., \& {Meiring}, J.~G. 2006, in Society of
  Photo-Optical Instrumentation Engineers (SPIE) Conference Series, Vol. 6267,
  Society of Photo-Optical Instrumentation Engineers (SPIE) Conference Series,
  ed. L.~M. {Stepp}, 62670Z

\bibitem[{{Burgh} {et~al.}(2003){Burgh}, {Nordsieck}, {Kobulnicky}, {Williams},
  {O'Donoghue}, {Smith}, \& {Percival}}]{burgh+03}
{Burgh}, E.~B., {Nordsieck}, K.~H., {Kobulnicky}, H.~A., {et~al.} 2003, in
  Society of Photo-Optical Instrumentation Engineers (SPIE) Conference Series,
  Vol. 4841, Instrument Design and Performance for Optical/Infrared
  Ground-based Telescopes, ed. M.~{Iye} \& A.~F.~M. {Moorwood}, 1463--1471

\bibitem[{{Clayton} \& {Osborne}(1994)}]{clayton+94}
{Clayton}, K.~L. \& {Osborne}, J.~P. 1994, \mnras, 268, 229

\bibitem[{{Covington} {et~al.}(2022){Covington}, {Shaw}, {Mukai},
  {Littlefield}, {Heinke}, {Plotkin}, {Barrett}, {Boardman}, {Boyd}, {Brincat},
  {Carstens}, {Collins}, {Cook}, {Cooney}, {Fern{\'a}ndez}, {Dufoer}, {Dvorak},
  {Galdies}, {Goff}, {Hambsch}, {Johnston}, {Jones}, {Menzies}, {Monard},
  {Morelle}, {Nelson}, {{\"O}gmen}, {Rock}, {Sabo}, {Seargeant}, {Stone},
  {Ulowetz}, \& {Vanmunster}}]{covington+22}
{Covington}, A.~E., {Shaw}, A.~W., {Mukai}, K., {et~al.} 2022, \apj, 928, 164

\bibitem[{{Creevey} {et~al.}(2022){Creevey}, {Sordo}, {Pailler}, {Fr{\'e}mat},
  {Heiter}, {Th{\'e}venin}, {Andrae}, {Fouesneau}, {Lobel}, {Bailer-Jones},
  {Garabato}, {Bellas-Velidis}, {Brugaletta}, {Lorca}, {Ordenovic}, {Palicio},
  {Sarro}, {Delchambre}, {Drimmel}, {Rybizki}, {Torralba Elipe}, {Korn},
  {Recio-Blanco}, {Schultheis}, {De Angeli}, {Montegriffo}, {Abreu Aramburu},
  {Accart}, {{\'A}lvarez}, {Bakker}, {Brouillet}, {Burlacu}, {Carballo},
  {Casamiquela}, {Chiavassa}, {Contursi}, {Cooper}, {Dafonte}, {Dapergolas},
  {de Laverny}, {Dharmawardena}, {Edvardsson}, {Le Fustec},
  {Garc{\'\i}a-Lario}, {Garc{\'\i}a-Torres}, {Gomez},
  {Gonz{\'a}lez-Santamar{\'\i}a}, {Hatzidimitriou}, {Jean-Antoine Piccolo},
  {Kontizas}, {Kordopatis}, {Lanzafame}, {Lebreton}, {Licata}, {Lindstr{\o}m},
  {Livanou}, {Magdaleno Romeo}, {Manteiga}, {Marocco}, {Marshall}, {Mary},
  {Nicolas}, {Pallas-Quintela}, {Panem}, {Pichon}, {Poggio}, {Riclet}, {Robin},
  {Santove{\~n}a}, {Silvelo}, {Slezak}, {Smart}, {Soubiran}, {S{\"u}veges},
  {Ulla}, {Utrilla}, {Vallenari}, {Zhao}, {Zorec}, {Barrado}, {Bijaoui},
  {Bouret}, {Blomme}, {Brott}, {Cassisi}, {Kochukhov}, {Martayan}, {Shulyak},
  \& {Silvester}}]{creevey+22}
{Creevey}, O.~L., {Sordo}, R., {Pailler}, F., {et~al.} 2022, arXiv e-prints,
  arXiv:2206.05864

\bibitem[{{Dorman} \& {Arnaud}(2001)}]{dorman+01}
{Dorman}, B. \& {Arnaud}, K.~A. 2001, in Astronomical Society of the Pacific
  Conference Series, Vol. 238, Astronomical Data Analysis Software and Systems
  X, ed. J.~{Harnden}, F.~R., F.~A. {Primini}, \& H.~E. {Payne}, 415

\bibitem[{{Eyer} {et~al.}(2022){Eyer}, {Audard}, {Holl}, {Rimoldini},
  {Carnerero}, {Clementini}, {De Ridder}, {Distefano}, {Evans}, {Gavras},
  {Gomel}, {Lebzelter}, {Marton}, {Mowlavi}, {Panahi}, {Ripepi}, {Wyrzykowski},
  {Nienartowicz}, {Jevardat de Fombelle}, {Lecoeur-Taibi}, {Rohrbasser},
  {Riello}, {Garcia-Lario}, {Lanzafame}, {Mazeh}, {Raiteri}, {Zucker},
  {Abraham}, {Aerts}, {Aguado}, {Anderson}, {Bashi}, {Binnenfeld}, {Faigler},
  {Garofalo}, {Karbevska}, {Kospal}, {Kruszynska}, {Kun}, {Lanza}, {Leccia},
  {Marconi}, {Messina}, {Molinaro}, {Molnar}, {Muraveva}, {Musella}, {Nagy},
  {Pagano}, {Palaversa}, {Plachy}, {Rybicki}, {Shahaf}, {Szabados},
  {Szegedi-Elek}, {Trabucchi}, {Barblan}, \& {Roelens}}]{eyer+22}
{Eyer}, L., {Audard}, M., {Holl}, B., {et~al.} 2022, arXiv e-prints,
  arXiv:2206.06416

\bibitem[{{Feinstein} {et~al.}(2019){Feinstein}, {Montet}, {Foreman-Mackey},
  {Bedell}, {Saunders}, {Bean}, {Christiansen}, {Hedges}, {Luger}, {Scolnic},
  \& {Cardoso}}]{feinstein+19}
{Feinstein}, A.~D., {Montet}, B.~T., {Foreman-Mackey}, D., {et~al.} 2019,
  \pasp, 131, 094502

\bibitem[{{Ferrario} {et~al.}(1989){Ferrario}, {Wickramasinghe}, \&
  {Tuohy}}]{ferrario+89}
{Ferrario}, L., {Wickramasinghe}, D.~T., \& {Tuohy}, I.~R. 1989, \apj, 341, 327

\bibitem[{{Fouesneau} {et~al.}(2022){Fouesneau}, {Fr{\'e}mat}, {Andrae},
  {Korn}, {Soubiran}, {Kordopatis}, {Vallenari}, {Heiter}, {Creevey}, {Sarro},
  {de Laverny}, {Lanzafame}, {Lobel}, {Sordo}, {Rybizki}, {Slezak},
  {{\'A}lvarez}, {Drimmel}, {Garabato}, {Delchambre}, {Bailer-Jones},
  {Hatzidimitriou}, {Lorca}, {Le Fustec}, {Pailler}, {Mary}, {Robin},
  {Utrilla}, {Abreu Aramburu}, {Bakker}, {Bellas-Velidis}, {Bijaoui}, {Blomme},
  {Bouret}, {Brouillet}, {Brugaletta}, {Burlacu}, {Carballo}, {Casamiquela},
  {Chaoul}, {Chiavassa}, {Contursi}, {Cooper}, {Dafonte}, {Demouchy},
  {Dharmawardena}, {Garc{\'\i}a-Lario}, {Garc{\'\i}a-Torres}, {Gomez},
  {Gonz{\'a}lez-Santamar{\'\i}a}, {Jean-Antoine Piccolo}, {Kontizas},
  {Lebreton}, {Licata}, {Lindstr{\o}m}, {Livanou}, {Magdaleno Romeo},
  {Manteiga}, {Marocco}, {Martayan}, {Marshall}, {Nicolas}, {Ordenovic},
  {Palicio}, {Pallas-Quintela}, {Pichon}, {Poggio}, {Recio-Blanco}, {Riclet},
  {Santove{\~n}a}, {Schultheis}, {Segol}, {Silvelo}, {Smart}, {S{\"u}veges},
  {Th{\'e}venin}, {Torralba Elipe}, {Ulla}, {van Dillen}, {Zhao}, \&
  {Zorec}}]{fouesneau+22}
{Fouesneau}, M., {Fr{\'e}mat}, Y., {Andrae}, R., {et~al.} 2022, arXiv e-prints,
  arXiv:2206.05992

\bibitem[{{Frank} {et~al.}(1988){Frank}, {King}, \& {Lasota}}]{frank+88}
{Frank}, J., {King}, A.~R., \& {Lasota}, J.~P. 1988, \aap, 193, 113

\bibitem[{{Freyberg} {et~al.}(2020){Freyberg}, {Perinati}, {Pacaud}, {Eraerds},
  {Churazov}, {Dennerl}, {Predehl}, {Merloni}, {Meidinger}, {Bulbul},
  {Friedrich}, {Gilfanov}, {Tenzer}, {Pommranz}, {Eckert}, {Schmitt}, {Brusa},
  \& {Santangelo}}]{freyberg+20}
{Freyberg}, M., {Perinati}, E., {Pacaud}, F., {et~al.} 2020, in Society of
  Photo-Optical Instrumentation Engineers (SPIE) Conference Series, Vol. 11444,
  Society of Photo-Optical Instrumentation Engineers (SPIE) Conference Series,
  114441O

\bibitem[{{Gaia Collaboration} {et~al.}(2021){Gaia Collaboration}, {Brown},
  {Vallenari}, {Prusti}, {de Bruijne}, {Babusiaux}, {Biermann}, {Creevey},
  {Evans}, {Eyer}, \& et~al.}]{gaia+21}
{Gaia Collaboration}, {Brown}, A.~G.~A., {Vallenari}, A., {et~al.} 2021, \aap,
  649, A1

\bibitem[{{Hessman} {et~al.}(2000){Hessman}, {G{\"a}nsicke}, \&
  {Mattei}}]{hessman+00}
{Hessman}, F.~V., {G{\"a}nsicke}, B.~T., \& {Mattei}, J.~A. 2000, \aap, 361,
  952

\bibitem[{{Hill} {et~al.}(2022){Hill}, {Littlefield}, {Garnavich}, {Scaringi},
  {Szkody}, {Mason}, {Kennedy}, {Shaw}, \& {Covington}}]{hill+22}
{Hill}, K.~L., {Littlefield}, C., {Garnavich}, P., {et~al.} 2022, \aj, 163, 246

\bibitem[{{Kafka} \& {Honeycutt}(2005)}]{kafka+05}
{Kafka}, S. \& {Honeycutt}, R.~K. 2005, \aj, 130, 742

\bibitem[{{Kennedy} {et~al.}(2020){Kennedy}, {Garnavich}, {Littlefield},
  {Marsh}, {Callanan}, {Breton}, {Augusteijn}, {Wagner}, {Ashley}, \&
  {Neric}}]{kennedy+20}
{Kennedy}, M.~R., {Garnavich}, P.~M., {Littlefield}, C., {et~al.} 2020, \mnras,
  495, 4445

\bibitem[{{Knigge}(2006)}]{knigge+06}
{Knigge}, C. 2006, \mnras, 373, 484

\bibitem[{{Knigge} {et~al.}(2011){Knigge}, {Baraffe}, \&
  {Patterson}}]{knigge+11}
{Knigge}, C., {Baraffe}, I., \& {Patterson}, J. 2011, \apjs, 194, 28

\bibitem[{{Kraft} {et~al.}(1991){Kraft}, {Burrows}, \&
  {Nousek}}]{1991ApJ...374..344K}
{Kraft}, R.~P., {Burrows}, D.~N., \& {Nousek}, J.~A. 1991, \apj, 374, 344

\bibitem[{{Kuulkers} {et~al.}(2010){Kuulkers}, {Norton}, {Schwope}, \&
  {Warner}}]{kuulkers+10}
{Kuulkers}, E., {Norton}, A., {Schwope}, A., \& {Warner}, B. 2010, in Compact
  Stellar X-ray Sources, 421

\bibitem[{{Lamb} \& {Masters}(1979)}]{lamb+79}
{Lamb}, D.~Q. \& {Masters}, A.~R. 1979, \apjl, 234, L117

\bibitem[{{Liedahl} {et~al.}(1995){Liedahl}, {Osterheld}, \&
  {Goldstein}}]{liedahl+95}
{Liedahl}, D.~A., {Osterheld}, A.~L., \& {Goldstein}, W.~H. 1995, \apjl, 438,
  L115

\bibitem[{{Lightkurve Collaboration} {et~al.}(2018){Lightkurve Collaboration},
  {Cardoso}, {Hedges}, {Gully-Santiago}, {Saunders}, {Cody}, {Barclay}, {Hall},
  {Sagear}, {Turtelboom}, {Zhang}, {Tzanidakis}, {Mighell}, {Coughlin}, {Bell},
  {Berta-Thompson}, {Williams}, {Dotson}, \& {Barentsen}}]{lcsoftware2018}
{Lightkurve Collaboration}, {Cardoso}, J.~V.~d.~M., {Hedges}, C., {et~al.}
  2018, {Lightkurve: Kepler and TESS time series analysis in Python},
  Astrophysics Source Code Library

\bibitem[{{Littlefield} {et~al.}(2020){Littlefield}, {Garnavich}, {Kennedy},
  {Patterson}, {Kemp}, {Stiller}, {Hambsch}, {Heras}, {Myers}, {Stone},
  {Sj{\"o}berg}, {Dvorak}, {Nelson}, {Popov}, {Bonnardeau}, {Vanmunster}, {de
  Miguel}, {Alton}, {Harris}, {Cook}, {Graham}, {Brincat}, {Lane}, {Foster},
  {Pickard}, {Sabo}, {Vietje}, {Lemay}, {Briol}, {Krumm}, {Dadighat}, {Goff},
  {Solomon}, {Padovan}, {Bolt}, {Kardasis}, {Deback{\`e}re}, {Thrush}, {Stein},
  {Walter}, {Coulter}, {Tsehmeystrenko}, {Gout}, {Lewin}, {Galdies},
  {Fernandez}, {Walker}, {Boardman}, \& {Pellett}}]{littlefield+20}
{Littlefield}, C., {Garnavich}, P., {Kennedy}, M.~R., {et~al.} 2020, \apj, 896,
  116

\bibitem[{{Livio} \& {Pringle}(1994)}]{livio+94}
{Livio}, M. \& {Pringle}, J.~E. 1994, \apj, 427, 956

\bibitem[{{Lomb}(1976)}]{lomb+76}
{Lomb}, N.~R. 1976, \apss, 39, 447

\bibitem[{{Markley} \& {Crassidis}(2014)}]{markley14}
{Markley}, F.~L. \& {Crassidis}, J.~L. 2014, {Fundamentals of Spacecraft
  Attitude Determination and Control}

\bibitem[{{Mewe} {et~al.}(1985){Mewe}, {Gronenschild}, \& {van den
  Oord}}]{mewe+85}
{Mewe}, R., {Gronenschild}, E.~H.~B.~M., \& {van den Oord}, G.~H.~J. 1985,
  \aaps, 62, 197

\bibitem[{{Mukai}(2017)}]{mukai17}
{Mukai}, K. 2017, \pasp, 129, 062001

\bibitem[{{Nauenberg}(1972)}]{nauenberg+72}
{Nauenberg}, M. 1972, \apj, 175, 417

\bibitem[{{Norton} {et~al.}(1997){Norton}, {Hellier}, {Beardmore}, {Wheatley},
  {Osborne}, \& {Taylor}}]{norton+97}
{Norton}, A.~J., {Hellier}, C., {Beardmore}, A.~P., {et~al.} 1997, \mnras, 289,
  362

\bibitem[{{Oke}(1990)}]{oke90}
{Oke}, J.~B. 1990, \aj, 99, 1621

\bibitem[{{Pala} {et~al.}(2021){Pala}, {G{\"a}nsicke}, {Belloni}, {Parsons},
  {Marsh}, {Schreiber}, {Breedt}, {Knigge}, {Sion}, {Szkody}, {Townsley},
  {Bildsten}, {Boyd}, {Cook}, {De Martino}, {Godon}, {Kafka}, {Kouprianov},
  {Long}, {Monard}, {Myers}, {Nelson}, {Nogami}, {Oksanen}, {Pickard},
  {Poyner}, {Reichart}, {Rodriguez Perez}, {Shears}, {Stubbings}, \&
  {Toloza}}]{pala+21}
{Pala}, A.~F., {G{\"a}nsicke}, B.~T., {Belloni}, D., {et~al.} 2021, \mnras
  [\eprint[arXiv]{2111.13706}]

\bibitem[{{Predehl} {et~al.}(2021){Predehl}, {Andritschke}, {Arefiev},
  {Babyshkin}, {Batanov}, {Becker}, {B{\"o}hringer}, {Bogomolov}, {Boller},
  {Borm}, {Bornemann}, {Br{\"a}uninger}, {Br{\"u}ggen}, {Brunner}, {Brusa},
  {Bulbul}, {Buntov}, {Burwitz}, {Burkert}, {Clerc}, {Churazov}, {Coutinho},
  {Dauser}, {Dennerl}, {Doroshenko}, {Eder}, {Emberger}, {Eraerds},
  {Finoguenov}, {Freyberg}, {Friedrich}, {Friedrich}, {F{\"u}rmetz},
  {Georgakakis}, {Gilfanov}, {Granato}, {Grossberger}, {Gueguen}, {Gureev},
  {Haberl}, {H{\"a}lker}, {Hartner}, {Hasinger}, {Huber}, {Ji}, {Kienlin},
  {Kink}, {Korotkov}, {Kreykenbohm}, {Lamer}, {Lomakin}, {Lapshov}, {Liu},
  {Maitra}, {Meidinger}, {Menz}, {Merloni}, {Mernik}, {Mican}, {Mohr},
  {M{\"u}ller}, {Nandra}, {Nazarov}, {Pacaud}, {Pavlinsky}, {Perinati},
  {Pfeffermann}, {Pietschner}, {Ramos-Ceja}, {Rau}, {Reiffers}, {Reiprich},
  {Robrade}, {Salvato}, {Sanders}, {Santangelo}, {Sasaki}, {Scheuerle},
  {Schmid}, {Schmitt}, {Schwope}, {Shirshakov}, {Steinmetz}, {Stewart},
  {Str{\"u}der}, {Sunyaev}, {Tenzer}, {Tiedemann}, {Tr{\"u}mper}, {Voron},
  {Weber}, {Wilms}, \& {Yaroshenko}}]{predehl+21}
{Predehl}, P., {Andritschke}, R., {Arefiev}, V., {et~al.} 2021, \aap, 647, A1

\bibitem[{{Ramsay} \& {Cropper}(2004)}]{ramsay+04}
{Ramsay}, G. \& {Cropper}, M. 2004, \mnras, 347, 497

\bibitem[{{Ramsay} {et~al.}(2009){Ramsay}, {Rosen}, {Hakala}, \&
  {Barclay}}]{ramsay+09}
{Ramsay}, G., {Rosen}, S., {Hakala}, P., \& {Barclay}, T. 2009, \mnras, 395,
  416

\bibitem[{{Ricker} {et~al.}(2014){Ricker}, {Winn}, {Vanderspek}, {Latham},
  {Bakos}, {Bean}, {Berta-Thompson}, {Brown}, {Buchhave}, {Butler}, {Butler},
  {Chaplin}, {Charbonneau}, {Christensen-Dalsgaard}, {Clampin}, {Deming},
  {Doty}, {De Lee}, {Dressing}, {Dunham}, {Endl}, {Fressin}, {Ge}, {Henning},
  {Holman}, {Howard}, {Ida}, {Jenkins}, {Jernigan}, {Johnson}, {Kaltenegger},
  {Kawai}, {Kjeldsen}, {Laughlin}, {Levine}, {Lin}, {Lissauer}, {MacQueen},
  {Marcy}, {McCullough}, {Morton}, {Narita}, {Paegert}, {Palle}, {Pepe},
  {Pepper}, {Quirrenbach}, {Rinehart}, {Sasselov}, {Sato}, {Seager},
  {Sozzetti}, {Stassun}, {Sullivan}, {Szentgyorgyi}, {Torres}, {Udry}, \&
  {Villasenor}}]{ricker+14}
{Ricker}, G.~R., {Winn}, J.~N., {Vanderspek}, R., {et~al.} 2014, in Society of
  Photo-Optical Instrumentation Engineers (SPIE) Conference Series, Vol. 9143,
  Space Telescopes and Instrumentation 2014: Optical, Infrared, and Millimeter
  Wave, ed. J.~{Oschmann}, Jacobus~M., M.~{Clampin}, G.~G. {Fazio}, \& H.~A.
  {MacEwen}, 914320

\bibitem[{{Ritter} \& {Kolb}(2003)}]{ritter+03}
{Ritter}, H. \& {Kolb}, U. 2003, \aap, 404, 301

\bibitem[{{Scargle}(1982)}]{scargle+82}
{Scargle}, J.~D. 1982, \apj, 263, 835

\bibitem[{{Schwarz} {et~al.}(2009){Schwarz}, {Schwope}, {Vogel}, {Dhillon},
  {Marsh}, {Copperwheat}, {Littlefair}, \& {Kanbach}}]{schwarz+09}
{Schwarz}, R., {Schwope}, A.~D., {Vogel}, J., {et~al.} 2009, \aap, 496, 833

\bibitem[{{Schwope} {et~al.}(2022){Schwope}, {Buckley}, {Malyali}, {Potter},
  {K{\"o}nig}, {Arcodia}, {Gromadzki}, \& {Rau}}]{schwope+22}
{Schwope}, A., {Buckley}, D. A.~H., {Malyali}, A., {et~al.} 2022, \aap, 661,
  A43

\bibitem[{{Schwope} {et~al.}(2008){Schwope}, {Vogel}, {Schwarz}, {Walter},
  {Burwitz}, \& {Reinsch}}]{schwope+08}
{Schwope}, A., {Vogel}, J., {Schwarz}, R., {et~al.} 2008, in The X-ray Universe
  2008, 73

\bibitem[{{Schwope} {et~al.}(2002){Schwope}, {Brunner}, {Hambaryan}, \&
  {Schwarz}}]{schwope+02}
{Schwope}, A.~D., {Brunner}, H., {Hambaryan}, V., \& {Schwarz}, R. 2002, in
  Astronomical Society of the Pacific Conference Series, Vol. 261, The Physics
  of Cataclysmic Variables and Related Objects, ed. B.~T. {G{\"a}nsicke},
  K.~{Beuermann}, \& K.~{Reinsch}, 102

\bibitem[{{Schwope} {et~al.}(2001){Schwope}, {Schwarz}, {Sirk}, \&
  {Howell}}]{schwope+01}
{Schwope}, A.~D., {Schwarz}, R., {Sirk}, M., \& {Howell}, S.~B. 2001, \aap,
  375, 419

\bibitem[{{Schwope} {et~al.}(2020){Schwope}, {Worpel}, {Traulsen}, \&
  {Sablowski}}]{schwope+20}
{Schwope}, A.~D., {Worpel}, H., {Traulsen}, I., \& {Sablowski}, D. 2020, \aap,
  642, A134

\bibitem[{{Sunyaev} {et~al.}(2021){Sunyaev}, {Arefiev}, {Babyshkin},
  {Bogomolov}, {Borisov}, {Buntov}, {Brunner}, {Burenin}, {Churazov},
  {Coutinho}, {Eder}, {Eismont}, {Freyberg}, {Gilfanov}, {Gureyev}, {Hasinger},
  {Khabibullin}, {Kolmykov}, {Komovkin}, {Krivonos}, {Lapshov}, {Levin},
  {Lomakin}, {Lutovinov}, {Medvedev}, {Merloni}, {Mernik}, {Mikhailov},
  {Molodtsov}, {Mzhelsky}, {M{\"u}ller}, {Nandra}, {Nazarov}, {Pavlinsky},
  {Poghodin}, {Predehl}, {Robrade}, {Sazonov}, {Scheuerle}, {Shirshakov},
  {Tkachenko}, \& {Voron}}]{sunyaev+21}
{Sunyaev}, R., {Arefiev}, V., {Babyshkin}, V., {et~al.} 2021, \aap, 656, A132

\bibitem[{{Szkody}(1998)}]{szkody98}
{Szkody}, P. 1998, in Astronomical Society of the Pacific Conference Series,
  Vol. 137, Wild Stars in the Old West, ed. S.~{Howell}, E.~{Kuulkers}, \&
  C.~{Woodward}, 18

\bibitem[{{Terndrup} {et~al.}(2002){Terndrup}, {Pinsonneault}, {Jeffries},
  {Ford}, {Stauffer}, \& {Sills}}]{terndrup+02}
{Terndrup}, D.~M., {Pinsonneault}, M., {Jeffries}, R.~D., {et~al.} 2002, \apj,
  576, 950

\bibitem[{{Tody}(1986)}]{tody86}
{Tody}, D. 1986, in Society of Photo-Optical Instrumentation Engineers (SPIE)
  Conference Series, Vol. 627, Instrumentation in astronomy VI, ed. D.~L.
  {Crawford}, 733

\bibitem[{{VanderPlas}(2018)}]{vanderplas+18}
{VanderPlas}, J.~T. 2018, \apjs, 236, 16

\bibitem[{{Vogel} {et~al.}(2008){Vogel}, {Byckling}, {Schwope}, {Osborne},
  {Schwarz}, \& {Watson}}]{vogel+08}
{Vogel}, J., {Byckling}, K., {Schwope}, A., {et~al.} 2008, \aap, 485, 787

\bibitem[{{Wahba}(1965)}]{wahba}
{Wahba}, G. 1965, SIAM Review, 7, 409

\bibitem[{{Warner}(1995)}]{warner+95}
{Warner}, B. 1995, {Cataclysmic variable stars}, Vol.~28

\bibitem[{{Warren} {et~al.}(1995){Warren}, {Sirk}, \& {Vallerga}}]{warren+95}
{Warren}, J.~K., {Sirk}, M.~M., \& {Vallerga}, J.~V. 1995, \apj, 445, 909

\bibitem[{{Watson} {et~al.}(1989){Watson}, {King}, {Jones}, \&
  {Motch}}]{watson+89}
{Watson}, M.~G., {King}, A.~R., {Jones}, M.~H., \& {Motch}, C. 1989, \mnras,
  237, 299

\bibitem[{{Webb} {et~al.}(2018){Webb}, {Schwope}, {Zolotukhin}, {Lin}, \&
  {Rosen}}]{webb+18}
{Webb}, N.~A., {Schwope}, A., {Zolotukhin}, I., {Lin}, D., \& {Rosen}, S.~R.
  2018, \aap, 615, A133

\bibitem[{{Wilms} {et~al.}(2000){Wilms}, {Allen}, \& {McCray}}]{wilms+00}
{Wilms}, J., {Allen}, A., \& {McCray}, R. 2000, \apj, 542, 914

\bibitem[{{Wu} \& {Kiss}(2008)}]{wu+08}
{Wu}, K. \& {Kiss}, L.~L. 2008, \aap, 481, 433

\end{thebibliography}
\bibliographystyle{aa}

\end{document}